\documentstyle[pre,aps]{revtex}
\begin{document}
\title{Instability, Intermittency and Multiscaling in Discrete Growth Models 
of Kinetic Roughening}
\author{C. Dasgupta \cite{cd}, J. M. Kim \cite{jmk}, M. Dutta and S. Das Sarma}
\address{Department of Physics, University of Maryland, College
Park, Maryland 20742-4111.}
\date{\today}
\maketitle
\draft
\begin{abstract} 
We show by numerical simulations that discretized versions of commonly
studied continuum nonlinear growth
equations (such as the Kardar-Parisi-Zhang equation and the Lai-Das Sarma 
equation) and related atomistic models of epitaxial growth 
have a generic instability in which isolated pillars
(or grooves) on an otherwise flat interface grow in time when their height
(or depth) exceeds a critical value. Depending on the details of the model,
the instability found in the discretized
version may or may not 
be present in the truly continuum growth equation, indicating
that the behavior of discretized nonlinear growth equations may be
very different from that of their continuum counterparts. This
instability can be controlled either by the introduction of higher-order 
nonlinear terms with appropriate coefficients or by restricting the growth
of pillars (or grooves) by other means. A number of such ``controlled
instability'' models are studied by simulation. For appropriate choice of
the parameters used for controlling the instability, these models exhibit
intermittent behavior, characterized by multiexponent scaling of
height fluctuations, over the time interval during which the instability is
active. The behavior found in this regime is very 
similar to the ``turbulent'' behavior observed in recent simulations of
several one- and two-dimensional atomistic models of epitaxial growth.

\end{abstract}

\section{Introduction}
In recent years, much attention has been focused on the problem of
kinetic surface roughening associated with the nonequilibrium
dynamics of growing interfaces \cite{rev1,rev2}. A number of simple models of
epitaxial growth have been proposed and studied
\cite{kpz,lds,villain,dt,kds,krug,dlkg} analytically and
numerically, revealing a rich variety of
interesting phenomena. One such phenomenon for which no
explanation is currently available is
the multi-exponent scaling (``multiscaling'' in short)
of height fluctuations
found \cite{krug} in recent simulations \cite{krug,dlkg,bdk}
of a class of one-dimensional (1d) limited-mobility models of
epitaxial growth. This phenomenon is particularly interesting
because it exhibits a striking similarity \cite{krug} to the
intermittent multiscaling of velocity fluctuations in fully developed fluid
turbulence \cite{turb}. 

This paper describes the results of a numerical investigation of the 
origin of this interesting multiscaling behavior. Our study shows that
the multiscaling found in these models is closely related to an 
instability of discretized versions of commonly studied nonlinear
growth equations. In this instability, isolated structures (such as
pillars or grooves) on a flat interface tend to grow in time if
the ``size'' of the structure (i.e. height of a pillar or depth
of a groove) exceeds a critical value. We show that this 
instability is the cause of numerical difficulties encountered 
in earlier work \cite{tu,mkw} on numerical integration of discretized growth
equations. These difficulties were usually attributed to ``numerical artifacts''
in previous studies. In contrast, we show that these numerical difficulties
are the consequence of a genuine instability intrinsic to the discretized
growth equations.
This instability is found to be ``generic'' to a large class of
discretized growth equations with nonlinearities. In particular,
we find that this instability is present in 1d and 2d
versions of the conserved fourth-order growth equation
introduced by Lai and Das Sarma (LD) \cite{lds} and by Villain
\cite{villain}, and also in the 1d
Kardar-Parisi-Zhang (KPZ) equation \cite{kpz} with or without
noise. Since the 1d continuum KPZ equation without noise, which is exactly
solvable via a mapping to the diffusion equation by a Cole-Hopf
transformation \cite{rev2},
 {\em does not} have any instability, our results lead
to the important conclusion that the behavior of discretized nonlinear
growth equations may be very different from that of the corresponding
truly continuum
versions. Whether this instability
occurs or not during the time evolution of a system  started
from a flat initial state is determined by the nature of the
dynamic scaling exhibited by the system. We find that this
instability is inevitable at sufficiently long times in models
which exhibit ``anomalous'' dynamic scaling \cite {dgk} 
(provided the system size is sufficiently large to prevent
saturation), whereas models exhibiting conventional scaling show
this instability only if the value of a dimensionless coupling
constant (defined below in terms of the values of the parameters in the
original growth equation and the length scale of discretization)
exceeds a non-zero critical value. A similar instability is found in 
an atomistic model \cite{kds} which is believed to provide an exact discrete
representation of the continuum LD growth equation.

Next, we show that this
instability can be controlled by introducing
higher-order nonlinear terms with appropriate coefficients.
These higher-order terms cut off the growth of pillars/grooves
at large values of the height/depth. The instability in atomistic growth
models can also be 
controlled by modifying the deposition rule in a way that restricts the 
growth of the height of a pillar or the depth of a groove.
We find that such models with controlled instability
exhibit deviations from simple scaling in the time interval during
which the instability is operative. If the parameter(s) used in 
the control of the instability is (are) chosen properly, then 
the behavior in this regime is found to be very similar to
the multiscaling observed in simulations
\cite{krug,dlkg,bdk} of atomistic growth models. The exponents 
which describe this approximate multiscaling behavior are 
non-universal; their values depend on the way the instability is
controlled. The overall picture that emerges from this study is qualitatively
similar to that suggested in the analytic work of Ref. 
\cite{bdk}. In particular, our work suggests that the
multiscaling behavior observed \cite{krug} 
in the 1d Das Sarma-Tamborenea (DT) model is
described by the LD equation supplemented by a set of
higher-order nonlinear terms with appropriate coefficients. 

As noted in Ref.\cite{krug}, the multiscaling
found in simulations of growth models is very similar to the
intermittent multifractal behavior observed in fluid turbulence. It is
interesting to note in this context that our proposed
explanation of multiscaling in growth models
is conceptually and formally 
similar to a recent proposal \cite{ep} which suggests 
that the multiscaling of structure functions in turbulence may
be explained in terms of singularities occurring on a dense set
of space-time points.

The rest of this paper is organized as follows. In section II, we define 
the models considered in our study and the various correlation functions
measured in our simulations to probe multiscaling behavior. Section III
contains a detailed account of the instability we find in discretized
growth equations and in an atomistic growth model. In section IV, we describe
the results of simulations of models in which the instability is controlled.
The behavior found in these simulations is compared and contrasted with
the multiscaling behavior observed in previous simulations of the DT and
related models. Section V contains a a summary of our findings and a
discussion of the implications of our results. A short paper describing
the main results of our study has been submitted for publication \cite{prl}.

\section {Models and Definitions}
Our work involves detailed numerical studies of
two continuum growth equations, namely the LD equation and the
KPZ equation. These equations are studied using direct numerical
integration. We have also studied by numerical simulations an
atomistic version of the LD equation introduced by Kim and Das
Sarma \cite{kds}. The LD equation we consider has the form
\begin{equation}
\partial h^{\prime}({\bf r},t)/\partial t = -\nu \nabla^4
h^{\prime} + \lambda_1 \nabla^2 |{\bf \nabla}h^{\prime}|^2 + 
\eta({\bf r},t), \label{eqn1}
\end{equation}
where $h^{\prime}({\bf r},t)$ represents the ``height'' variable at the
point ${\bf r}$ at time $t$, ${\bf \nabla}$ and $\nabla ^2$
represent, respectively, the spatial derivative and Laplacian
operators in $d$ dimensions (the dimension of the substrate),
and $\eta$ is a Gaussian random noise with correlations
\begin{equation}
<\eta({\bf r},t) \eta({\bf r^{\prime}}, t^{\prime})> = 2 D \delta({\bf r} -
{\bf r^{\prime}}) \delta (t - t^{\prime}). \label{eqn2}
\end{equation}
The KPZ equation in $(d+1)$ dimensions has the form
\begin{equation}
\partial h^{\prime}({\bf r},t)/\partial t = \nu \nabla^2
h^{\prime} + \lambda_1 |{\bf \nabla} h^{\prime}|^2 + \eta({\bf
r},t). \label{eqn3}
\end{equation}
We numerically integrate these two equations using a simple
Euler scheme \cite{tu,kds1}. To this end, we first define dimensionless
variables 
\begin{equation}
{\bf x} \equiv {\bf r}/r_0,\,\, \tau \equiv t/t_0,\,\, h \equiv
h^{\prime}/h_0, \label{eqn4} 
\end{equation}
where $r_0,\, t_0$ and $h_0$ are appropriately chosen units of
length, time and height, respectively. We then discretize in
space and time by defining the dimensionless discretization scale
$\Delta x$ and the integration time step
$\Delta \tau$. Using a proper choice of the units $ t_0$ and
$h_0$, the equations (\ref{eqn1}) and (\ref{eqn3}) can then
be represented by the following two update schemes:
\begin{equation}
h_i(\tau + \Delta \tau) - h_i(\tau) = \Delta
\tau \tilde{\nabla}^{2} [ - \tilde{\nabla}^{2} h_i(
\tau) + 
\lambda |\tilde{{\bf \nabla}} h_i(\tau)|^2] +\sqrt{\Delta
\tau} \eta_i^{\prime}(\tau), \label{eqn5}
\end{equation}
and
\begin{equation}
h_i(\tau + \Delta \tau) - h_i(\tau) = \Delta
\tau [ \tilde{\nabla}^{2} h_i(
\tau) +
\lambda |\tilde{{\bf \nabla}} h_i(\tau)|^2] +\sqrt{\Delta
\tau} \eta_i^{\prime}(\tau). \label{eqn6}
\end{equation}
In these equations, $h_i(\tau) \equiv h({\bf x}_i, \tau)$ represents the
dimensionless height variable at the lattice point $i$ at
dimensionless time $\tau$, $\tilde{\bf \nabla}$ and
$\tilde{\nabla}^{2}$ are lattice versions of the derivative
and Laplacian operators, and $\eta_i^{\prime}(\tau)$ is a random variable with
zero average and variance equal to unity. In most of our
calculations, we use the following definitions for the lattice
derivatives: 
\begin{eqnarray}
\tilde{\nabla}_j f_i &=& 0.5 [f({\bf x}_i + {\bf j}
\Delta x) - f({\bf x}_i - {\bf j} \Delta x)], \label{eqn7}\\
\tilde{\nabla}^{2}_j f_i &=& f({\bf x}_i + {\bf j}
\Delta x) + f({\bf x}_i - {\bf j} \Delta x) -2 f({\bf x}_i),
\label{eqn8} 
\end{eqnarray}
where $\bf j$ is an unit vector in the $j$th direction. In some
of our calculations, we have also used a more accurate
representation \cite{as} of the lattice derivatives involving
two neighbors on each side. The dimensionless parameter
$\lambda$ appearing in Eqs (\ref{eqn5}) and
(\ref{eqn6}) has the form
\begin{equation}
\lambda = \sqrt{2} (a_0/l_0)^{(4-d)/2}, \,\, l_0 \equiv
(\frac{\nu^3}{\lambda^2_1 D})^{1/(4-d)} \label{eqn9}
\end{equation}
for the LD equation and
\begin{equation}
\lambda = \sqrt{2} (a_0/l_0)^{(2-d)/2}, \,\, l_0 \equiv
(\frac{\nu^3}{\lambda^2_1 D})^{1/(2-d)} \label{eqn10}
\end{equation}
for the KPZ equation. In these equations, $a_0 \equiv r_0 \Delta x$
is the discretization scale (lattice spacing) and $l_0$ is
a characteristic length determined by the parameters
$\nu,\,\lambda_1$ and $D$ of the original
continuum growth equation. Note that in $d$ = 1, the value of
$\lambda$ would vanish in the true continuum limit,
$\Delta x \rightarrow 0$, for both LD and KPZ equations . However,
one should remember that a short-distance cutoff (perhaps of
atomic scale, e.g. the lattice spacing) 
is present in all physical situations, and it is
not legitimate to use a value of $\Delta x$ smaller than this
cutoff $a_{min}$. Therefore, the smallest value
($\lambda_{min}$) that the coupling
constant $\lambda$ can have is $\sqrt{2} (a_{min}/l_0)^{3/2}
$ and $\sqrt{2} (a_{min}/l_0)^{1/2} $ for the
1d LD and KPZ equations, respectively.

We have also studied an atomistic version \cite{kds} of the LD
equation in which the height variables $\{ h_i \}$ are integers. The time
evolution of this model is defined by the following deposition
rule. First, a site (say $i$) is chosen at random. Then the
quantity 
\begin{equation}
K_i(\{ h_j \}) =
-\tilde{\nabla}^{2}h_i + \lambda |\tilde{{\bf \nabla}}h_i|^2
\label{eqn11} 
\end{equation}
is calculated for the site $i$ and all its nearest 
neighbors. Then, a particle is added to the site that has the
smallest value of $K$ among the site $i$ and its nearest
neighbors. In the case of a tie for the smallest value, 
the site $i$ is chosen if it is
involved in the tie; otherwise, one of the
sites involved in the tie is chosen randomly. Note that this
model also involves only one dimensionless parameter, $\lambda$.
In this model, ``time'' is measured by the number of layers
deposited. We call this model the KD model \cite{kds} below.

The possibility of multiscaling was investigated  in our 
simulations by
monitoring different moments of the nearest-neighbor height
difference and the height difference correlation function. 
Following Ref. \onlinecite{krug}, we define
\begin{equation}
\sigma_q(\tau) \equiv <(s_i(\tau))^q>^{1/q}, \,\, s_i(t) = |h_{i+1}(\tau) -
h_i(\tau)|, \label{eqn12}
\end{equation}
and
\begin{equation}
G_q(l,\tau) \equiv <|h_{i+l}(\tau) - h_i(\tau)|^q>^{1/q}, \label{eqn13}
\end{equation}
where we have used the simplified notation $
h_{i+l} = h(x_i + l \Delta x)$ for the 1d system.
In these equations, the average $< \ldots >$ represents an
average over the site index $i$ and different runs corresponding
to different realizations of the random noise. Before saturation
(i.e. for $\tau \ll L^z$ where $L$ is the size of the system
and $z$ is the dynamical exponent), the quantities $\{ \sigma_q (\tau) \}$
are expected to show a  power-law growth in time $\tau$ in models 
which exhibit anomalous dynamic scaling:
\begin{equation}
\sigma_q(\tau) \approx \tau^{\alpha_q/z}. \label{eqn14}
\end{equation}
If the exponent $\alpha_q$ depends on the value of $q$, then the
model is said to exhibit multiscaling. Thus, whether
multiscaling is present or not can be easily tested by
monitoring the ratios $\sigma_q(\tau)/\sigma_1(\tau)$, $q$ = 2, 3 .. as
functions of time. The height-difference correlation functions
$G_q$ are expected to behave as
\begin{equation}
G_q(l,\tau) \approx |l|^{\zeta_q}, 1 \ll l \ll \xi(\tau) \approx
\tau^{1/z}. \label{eqn15}
\end{equation}
Again, multiscaling, characterized by a dependence of
the exponents $\zeta_q$ on $q$, can be tested by looking at the
$l$- dependence of the ratios $G_q(l)/G_1(l)$, $q$ = 2, 3 etc.
In our work, we consider the first four moments, $q$ = 1, 2, 3
and 4. We follow the notation of Ref.\cite{krug}, in contrast to the
notation of Ref.\cite{dlkg}, throughout this paper.

\section {Instability in Discrete Growth Equations}

In this section, we describe in detail the numerical calculations 
which lead to the conclusion that a generic instability is present in the
discretized growth equations defined in the preceding section.
We first studied the behavior of the 1d LD equation, Eq.(\ref{eqn5}),
for small values of the parameter $\lambda$ ($\lambda \le 2$).
In these runs, the system was started from a perfectly flat
state and its time evolution was simulated by integrating the
growth equations forward in time. 
Typical values of the parameters used in the simulation
are system size $L$ = 10$^3$, time step $\Delta \tau$ = 0.01 and
maximum time $\tau_{max}$ = 10$^4$. The results were averaged
over 10 - 50 independent runs.
In these runs for small values of $\lambda$, 
we find good agreement with the predictions of dynamical
renormalization group (DRG) calculations \cite{lds,dsk} and no
evidence of multiscaling. In particular, the exponent $\beta$
that describes the growth of the rms interface width $W$ with time is
found to have a value ($\simeq$ 0.34) which is close to the DRG result, $\beta
= 1/3$. We also find that the
exponents $\{\alpha_q/z\}$ are essentially independent of $q$ and
have a value close to zero (in the range 0.06 - 0.08), 
possibly indicating a logarithmic
growth in time. Typical results obtained for $\lambda = 1.0$ are shown
in Fig.1. As shown in the inset of this figure, the quantities $\{G_q(l)\}$
also do not show any indication of multiscaling. 
The exponents $\{\zeta_q\}$ have values in the
range 0.8 - 0.9, and are independent of $q$ within error bars. The
results obtained from simulations of the KD model for such
small values of $\lambda$ are very similar to those described
above. Results obtained for $\lambda = 0.5$ are shown in Fig.2. The exponents
calculated from the time-dependence of $W$  and $\sigma_q$ are, respectively,
$\beta \simeq 0.345$ and $\alpha_q/z \simeq 0.085$ for all $q$. The
behavior of the functions $\{G_q(l)\}$, shown in the inset of Fig.2, indicates
single-exponent scaling with $\zeta_q \simeq 0.9$ for all $q$.

The behavior observed for higher values of $\lambda$ is quite
different. In this case, the system exhibits the expected conventional
scaling behavior at short times. However, an apparent
``singularity'', indicated by a rapid growth of the height
variable, is found to occur at longer times. It is impossible to
follow numerically the evolution of the system beyond the time at which
this singularity occurs: attempts to do so lead to ``overflow''
on the computer. This instability was first observed by Tu
\cite{tu}; our results are quite similar to those
reported by him. The time at which this instability occurs shows
large run-to-run variations, with the average value
decreasing with increasing $\lambda$. A similar instability is
found in the KD model. Since the height variable in this
atomistic model can increase by only one unit at a time, there
is no divergence here. The instability in this model shows up as
a rapid increase of the interface width which corresponds to a
changeover from a power-law growth with an exponent close to 1/3 to a
linear growth in time. The results obtained from a simulation of the KD 
model with $\lambda = 1.0$ are shown in Fig.3. The occurrence of an instability
near $\tau = 100$ is clearly seen in the figure. 
It is interesting to note that the behavior of $W$ and $\{\sigma_q\}$ before
the occurrence of the instability is very similar to that found in simulations
for small values of $\lambda$ (see Fig.2). 
The occurrence of this instability was 
reported in Ref. \onlinecite{kds}. Thus, the observation of
these instabilities is not new. 
Our new results are about the origin of this instability, the apparently
``generic'' nature of this instability (in the sense that it appears to be
present in discretized versions of 
other commonly studied nonlinear growth equations such as the KPZ
equation), and the role it plays in the multiscaling phenomena observed 
\cite{krug,dlkg,bdk} in simulations of atomistic growth models.

We carried out a detailed investigation of the origin of this instability in
the discretized LD equation and the KD model and 
found that this instability is caused by the growth of
isolated structures, such as pillars and grooves, on a flat
interface. Either pillars or grooves are unstable in a
particular system; which one is unstable is determined by the
sign of $\lambda$. This asymmetry between pillars and grooves
results from the fact that the growth equations we consider are
not invariant under $h \rightarrow - h$. 
We find that pillars with heights exceeding a
certain ``critical'' value (which depends on the value of
$\lambda$, see below) grow in time in both the LD equation and the KD model
with positive $\lambda$. It is easy to show that in the absence of noise
($\eta^{\prime}$ = 0), an isolated pillar of height $h_0$ will initially
grow in time if $h_0 > 10/\lambda$. Consider an initial configuration
in which all sites except the central one have $h_i$ = 0 and the
central site has a height $h_0 > 0$ (a negative value of $h_0$ would
correspond to a groove at the center). The initial value of time derivative of 
the height at the central site is easily evaluated from Eq.(\ref{eqn5})
to be $-6h_0+\lambda h_0^2/2$. Similarly, the initial time derivative of the
height at one of nearest-neighboring sites of the central one is
obtained to be $4h_0 - \lambda h_0^2/2$. Clearly, the rate at which the 
difference between the heights at the central site and at one of its
nearest-neighboring sites initially changes with time is positive (i.e. the 
height of the pillar increases initially) if $h_0 > 10/\lambda$.
No analytic method is available for following the evolution of this state for
longer times or for taking into account the effects of the stochastic noise 
$\eta^{\prime}$. We therefore do this numerically and check at regular
time intervals whether the nearest-neighbor height
difference at the central site (defined as the larger of the two
height differences on the two sides) exceeds $h_0$ or not. By
repeating this procedure a large number of times, we are able to
calculate the probability $P(\tau)$ of the nearest-neighbor
height difference at the center exceeding the initial value
$h_0$ at a later time $\tau$. The results of such a study (for
$L$ =100, $\lambda$ = 1.0, $\Delta \tau$ = 0.01, 2000
independent runs) are shown in Fig.4. The probability of height
increase is found to be very close to zero for small values of
$h_0$. 
The growth probability 
begins to be non-zero as the value of $h_0$ exceeds $10/\lambda$. For
values of $h_0$ which are slightly higher than $10/\lambda$, the
growth probability is initially close to unity, but it
decreases rather quickly to zero (see the data for $h_0$ = 14
and 17 in Fig.4), indicating that the height
eventually decreases after an initial increase. The rate of the
initial growth of the height and the length of the time interval over
which the height remains greater than $h_0$ increase with $h_0$.
As $h_0$ is increased further, we encounter the instability
mentioned above. The height differences near the center grow
very rapidly, leading to overflow on the computer. To avoid this
problem, we stop the simulation of the time evolution when the
maximum value of the nearest-neighbor height difference exceeds a
preassigned cutoff value. This cutoff was chosen to be 1000 for the
data shown in Fig.4. The results are insensitive to the
value of this cutoff as long as it is large. In all runs stopped in
this way, the nearest-neighbor height difference at the central site 
is found to be larger than
$h_0$ when the run is stopped. 
These runs are counted as ones in which the nearest-neighbor 
height difference at
the center would remain greater than $h_0$ at later times. In fact, the large
value of the probability at $\tau$ = 1 for $h_0$ = 20 (see Fig.4)
arises exclusively from such runs. In other words, the height
difference at the center becomes smaller than $h_0$ within a
short time if the height differences do not exceed the cutoff value during the
time evolution of the system. This observation and the results of a rigorous 
analysis \cite{pbk} of the LD equation without noise suggest 
that the instability described above is not a true finite-time
singularity: the height of the pillar eventually decreases after
reaching a large but finite value. Here, we do not address the issue
of occurrence of a finite-time singularity in this model
because it is virtually impossible to determine numerically 
whether a true divergence of the height occurs or not. This
question is not crucial to our study: as described in section IV,
our main results 
are derived from models in which the growth of the height difference
is cut off at a finite value.

As shown in Fig.4, the probability of growth becomes
essentially independent of time near $\tau$ = 1. 
Fig.5 shows how the probability at $\tau$ = 1 depends on the
value of $h_0$. We show data obtained using three different values, 0.01,
0.001 and 0.0001, of the time step $\Delta \tau$. The observation that the
results obtained for these three very different values of $\Delta
\tau$ are nearly identical shows that
this instability is not a numerical artifact of 
not using a sufficiently small value
of the time step. From data of this kind, we define
a ``critical'' height $h_c$ for which the probability of growth
is 0.5. Our results indicate that the dependence of $h_c$ on
$\lambda$ is of the form
\begin{equation}
h_c(\lambda) \simeq A/\lambda \label{eqn16}
\end{equation}
with $A \simeq$ 20.0 for the LD equation. Our numerical results for
the values of $h_c$ are shown in Fig.6. 
The proportionality of $h_c$ to $1/\lambda$ may be understood from a simple 
dimensional argument. For a configuration in which the height variable is
$h_0$ at the central site and zero everywhere, the first term on the
right-hand side of Eq.(\ref{eqn5}), which tends to stabilize the system, is
proportional to $h_0$ at the central site and its nearest-neighbors. The
second term on the right-hand side of Eq.(\ref{eqn5}), which is the one
responsible for the instability, is proportional to $\lambda h_0^2$. It is,
therefore, obvious that the value of $h_0$ at which the destabilizing term
wins over the stabilizing one should be proportional to $1/\lambda$. The value
of the coefficient of proportionality $A$ is nontrivial and has to be 
determined numerically. 
We have also carried out
similar calculations using a more accurate, five-point
definition \cite{as} of the lattice derivatives. We find very
similar behavior, with a value of the parameter $A$
which is {\em smaller} than 20. This observation
indicates that the behavior described above is not an artifact
of using overly simple expressions for the lattice derivatives.

The development of the instability induced in the discretized 1d LD equation
by the presence of a high pillar in the initial state is illustrated in 
Fig.7 where we show the height profiles at times $\tau$ = 0.05, 0.1, 0.15
and 0.17, obtained by integrating the discretized LD equation ($L$ = 100,
$\lambda$ = 1.0, $\Delta \tau$ = 10$^{-4}$) from an initial state in which
the height is zero everywhere except at the central site where the height is
$h_0$ = 25. This value of $h_0$ is higher than the critical height $h_c$ for
the value of $\lambda$ used. As expected, the height of the pillar at the
center grows rapidly in time. At the same time, alternate grooves and pillars
form on both sides of the initial pillar and these grooves (pillars) become
higher (deeper) as time progresses. The formation of these grooves and
pillars is a consequence of the conservation law built into the LD equation.
In the run depicted in Fig.7, the maximum value of the nearest-neighbor 
height difference exceeded the cutoff of 1000 at time $\tau$ = 0.2.

Very similar results are obtained for the atomistic KD model.
A little algebra, similar to that described above, shows that in this model,
an attempt to deposit a ``particle'' at the site of a pillar of initial
height $h_0$ or at one of its nearest-neighboring sites leads to an increase
in the height of the pillar if $h_0 > 12/\lambda$. Our simulations (which
are exact because all variables in this model are discrete) show that
the height of a pillar continues to grow linearly in time if its initial
value is somewhat larger than $12/\lambda$. The development of the instability
in this model is very similar to that shown in Fig.7 for the discretized LD
equation.

The instability described above appears to be generic to
discretized versions of all commonly studied 
growth equations containing nonlinear terms. In
particular, we have found very similar results for two other
systems: the LD equation in (2+1) dimensions and the KPZ equation in
(1+1) dimensions. All the qualitative features of the
instability found in the 1d LD equation appear to be the present in two
dimensions.
Pillars of initial height $h_0$ become unstable in the 2d LD equation 
with positive $\lambda$ if $h_0 > h_c(\lambda)$. The dependence of $h_c$ on
$\lambda$ is well-described by 
$h_c(\lambda) \simeq A/\lambda$ with $A \simeq$
31 (see Fig.6).

The instability in the discretized KPZ equation, Eq.(\ref{eqn6}),
in one dimension with
$\lambda > 0$ is associated with grooves, not pillars. 
We have studied the 1d KPZ equation
with and without noise and found the instability to be very similar in the
two cases. The critical value of $h_0$ (the depth of an isolated groove) in 
the KPZ equation is
determined using a procedure similar to the one described above for the
LD equation. The instability criterion we use for the KPZ equation is slightly
different from the one described above. We define the probability of occurrence
of an instability as the ratio between the 
number of runs in which the maximum nearest-neighbor
height difference exceeds a preassigned cutoff value (taken to be 1000 in
our simulations) and the total number of runs. As noted above, this instability
criterion coincides with the criterion of the value of the nearest-neighbor 
height difference at the
central site exceeding $h_0$ in the LD equation. This is not so in the 1d
KPZ equation. In some of the runs, we find that the value of the
nearest-neighbor height difference at the central site is smaller than
$h_0$ when the maximum value
of the nearest-neighbor height difference reaches the cutoff. Evidently,
the presence of a deep groove in the initial state induces the formation of
large height differences at points which do not always coincide with the
initial location of the groove. The development of the instability in the
1d discretized KPZ equation is shown in Fig.8. The growth profiles shown for
times $\tau$ = 0.1, 0.3 and 0.5 are obtained for a $L$ = 100 system with
$\lambda$ = 1.0, using an integration time step $\Delta \tau = 10^{-4}$. The
initial state is one in which the height is zero everywhere except at the
central site where there is a groove of depth 30. As can be seen in Fig.8,
the groove at the center becomes deeper initially, but subsequently develops
into a ``mound'' with large values of the nearest-neighbor height difference
occurring at many points near the center. This is the reason why the maximum
value of the nearest-neighbor height difference does not always occur at the
central site in this system. A comparison of Fig.8 with Fig.7 clearly 
illustrates the important effects of a conservation law (which is present in
the LD equation, but absent in the KPZ equation) on the growth kinetics. The
value of the maximum nearest-neighbor height difference was found to exceed
the cutoff of 1000 at $\tau$ = 2.7 in the run for which the results are
shown in Fig.8. 

Typical results for the probability of
occurrence of an instability in the 1d KPZ equation ($\lambda$ = 1.0, $L$ =
100, $\Delta \tau$ =0.01, 2000 runs) are shown in Fig.9 for
three different values of $h_0$. 
As before, we define $h_c$ to be 
the value of $h_0$ at which the long-time value of the probability of
instability
reaches 0.5. The dependence of $h_c$ calculated in this way on the value of
$\lambda$ is shown in Fig.6. As expected, we find $h_c(\lambda) \simeq A/
\lambda$ with $A \simeq$ 25.0. The values of $h_c$ shown in Fig.6 were obtained
from numerical integrations using a time step $\Delta \tau$ =0.01. We have
repeated these calculations using smaller values of $\Delta \tau$. The 
observed dependence of the calculated value of $h_c$ on $\Delta \tau$ is 
more pronounced than that shown in Fig.5 for the 1d LD equation. 
However, any reasonable extrapolation of the results obtained for different
values of $\Delta \tau$ to the $\Delta \tau \rightarrow 0$ limit yields
results for $h_c$ which are not significantly different from 
those shown in Fig.6. We have
also checked directly the occurrence of the instability in the 1d 
discretized KPZ
equation without noise for initial conditions containing a deep groove 
for values of $\Delta \tau$ down to 10$^{-7}$.

Our conclusion about the existence of an instability in the discretized
version of the noiseless 1d
KPZ equation may appear surprising in view of the well-known fact that the
continuum KPZ equation without noise in one dimension {\em does not}
have any instability. The continuum equation can be mapped to a simple linear
diffusion equation by a Cole-Hopf transformation and solved exactly.
For any bounded initial condition, the asymptotic solution is one 
in which the height variable is constant. The absence of any instability in
the continuum equation, however, {\em does not necessarily} 
imply that the discretized
version, Eq.(\ref{eqn6}), should also be stable for any initial condition.
This is because
the application of a Cole-Hopf transformation to the discretized version
(cf. Eq.(\ref{eqn6})) of the KPZ equation {\em does not} reduce it 
to a discretized version of the
linear diffusion equation. The reason for this difference is simple: 
the algebra of derivatives does
not apply to difference operators if the nearest-neighbor height differences
are not vanishingly small. For this reason, the exact results available for 
the continuum equation do not in any way rule out the possibility of occurrence
of an instability in the discretized equation for initial conditions with
large nearest-neighbor height differences. Conclusions very similar to ours
about the occurrence of an instability in the 1d discretized noiseless KPZ
equation have recently been obtained independently by Newman and Bray \cite{nb}.

It is interesting to note that instabilities in numerical integrations of
the discretized KPZ equation in one and higher dimensions were noted in 
previous studies \cite{mkw}. These studies, however, attributed the observed 
instability to ``numerical artifacts'', with the implicit assumption that 
the instability would disappear if a sufficiently small value of the 
integration time step $\Delta \tau$ were used. Our work shows, for the
first time, that the instability found in these studies is an intrinsic
property of the discretized equation which {\em can not} be eliminated 
simply by using a sufficiently small value of $\Delta \tau$. The observation
of an instability in the discrete version of the noiseless KPZ equation in
one dimension brings out another important point which, to our knowledge,
has not been noted in the existing literature, namely, the behavior of
discretized versions of nonlinear growth equations may, under certain 
circumstances, be very different from the behavior of their continuum 
counterparts. These new observations have several important implications in
the study of growth equations. In particular, one important and inevitable
implication is that the discrete version of a nonlinear continuum growth
equation may, in principle, belong to a universality class which is different
from the universality class of the continuum equation.
A full discussion of these implications
is provided in section V.

So far, we have considered the time evolution of the discretized
growth equations from an
initial state in which a pillar or groove is present. A question
of obvious importance is whether such structures are
spontaneously generated during the evolution of the system from
a flat initial state. The answer to this question is
crucially related to the nature of dynamic scaling exhibited by
the model under consideration. In systems which exhibit normal
scaling, the nearest-neighbor height difference is not expected to
grow indefinitely in time; it should saturate quickly after an initial
growth. Such a system would spontaneously exhibit the
instability discussed above only if the value at which the
maximum nearest-neighbor height difference $s_{max}$
saturates is higher than (or at least, close to) the critical value, $h_c$,
defined above. Since $h_c$ decreases while the saturation value
of $s_{max}$ generally
increases with $\lambda$, we can define a  non-zero ``critical'' value,
$\lambda_c$, of $\lambda$ at which these two quantities become
equal. According to
the discussion above, systems with values of $\lambda$
substantially smaller than $\lambda_c$ are not expected to show
the instability during their time evolution from a flat state. 
In contrast, since  nearest-neighbor height
differences are expected to continue growing in time in models which exhibit
anomalous scaling, such systems should always show the
instability at sufficiently long times, provided the system size
is large enough to prevent saturation before the onset of
the instability. In other words, $\lambda_c$ is expected to be
zero for models with anomalous scaling. This conclusion is
different from that of Tu \cite{tu} who interpreted his
numerical results for the discretized 1d LD equation 
as evidence for the existence of a non-zero
$\lambda_c$ in this system. 

Our numerical results fully support these general
conclusions. In Fig.10, we show the time-dependence
of $s_{max}$, the maximum value of the nearest-neighbor height
difference averaged over a large number of runs starting
from a flat state. The system parameters are $L$ = 1000,
$\Delta \tau$ = 0.01, $\lambda$ = 4.0 for the 1d
LD equation; system size = 200$\times$200, $\Delta \tau $ = 0.01,
$\lambda$ = 5.0 for the 2d LD equation; and $L$ =
10$^4$, $\Delta \tau$ = 0.01, $\lambda$ = 5.0 for the
1d KPZ equation. A larger value of $L$ is needed 
for avoiding saturation in the KPZ equation because the value of
$z$ for this model is smaller. The quantity $s_{max}$ is
defined in the following way for the 2d system:
\begin{equation}
s_{max} \equiv max\{s_i\};\,\, 
s_i \equiv [(h({\bf x}_i + {\bf i}
\Delta x) - h({\bf x}_i))^2 
+ (h({\bf x}_i + {\bf j}
\Delta x) - h({\bf x}_i))^2]^{1/2}. \label{eqn17}
\end{equation}
The data shown were averaged over
200, 30 and 200 runs for the (1+1) LD, (2+1) LD and (1+1) 
KPZ equations, respectively. As
expected, $s_{max}$ saturates quickly for the last two models
which are expected
to exhibit normal scaling behavior. In contrast,
$s_{max}$ continues to grow in time in the 1d LD
equation which is expected to show anomalous scaling. The growth of
$s_{max}$ in time in this model is well-described by a
logarithmic form
\begin{equation}
s_{max}(\tau) \approx a + b \ln \tau, \label{eqn18}
\end{equation} 
where the parameter $b$ is numerically found to be proportional
to $\lambda$. This logarithmic dependence of $s_{max}$ on $\tau$
is consistent with predictions of dynamical renormalization
group calculations \cite{lds,dsk}. 

Fig.11 shows our numerical results for $\tau_{ins}$, the time
at which the instability occurs when the system evolves from a
flat initial state, for 1d LD and KPZ equations. We
operationally define $\tau_{ins}$ as the time at which $s_{max}$
reaches a cutoff value which is set at 1000. The instability
time shows very large run-to-run variations, and we find it more
appropriate to average $\ln \tau_{ins}$, rather than
$\tau_{ins}$ itself, over different runs. So, the data shown in
Fig.11 actually correspond to $\exp(<\ln \tau_{ins}>)$. The
data for the LD equation were obtained for $L$ = 1000, $\Delta \tau$
= 0.01, and averaged over 150, 200, 500, 1000 and 1000
independent runs for $\lambda$ = 3, 4, 5, 6 and 7, respectively.
From the results of Eqs (\ref{eqn16}) and (\ref{eqn18}), and the fact that the
coefficient $b$ of Eq. (\ref{eqn18}) is proportional to $\lambda$, it is
easy to show that the dependence of $\tau_{ins}$ on $\lambda$ in this
model should given by $\tau_{ins} \approx e^{B/\lambda^2}$. As
shown in Fig.11, this form does provide a good description of
the numerical data for small values of $\lambda$. The fact that
the values of $\tau_{ins}$ for $\lambda$ = 6 and 7 are lower
than those predicted by the fit to the data for smaller
$\lambda$ may be understood by noting that the initial growth of
$s_{max}$ with time is faster that that described by Eq.(\ref{eqn18}).

The data shown for the 1d KPZ equation were obtained
for $L$ = 10$^4$, $\Delta \tau$ = 0.01, and averaged over 200
runs. In 100 runs of length 10$^4$ units, we did not find any
occurrence of instability for $\lambda$ = 4. From the
observation that the distribution of $\tau_{ins}$ has a long
tail that extends to values much smaller than the average, we
can derive a conservative lower limit of 10$^6$ units for the 
average instability time for $\lambda$ = 4. This is indicated by
the arrow in Fig.11. These results strongly suggest that the
value of $\lambda_c$ lies between 4 and 5 in this system. Consequently,
the instability in this system may be avoided by choosing a 
small discretization scale $\Delta x$, one which would make the
value of $\lambda$ smaller than $\lambda_c$. However, as noted
before, the value of $\lambda$ in a real system can not be made arbitrarily
small, and the instability can not be avoided if the ``bare''
parameters are such that $\lambda_{min} > \lambda_c$. 

It should be mentioned that the above discussion about the possibility of
occurrence of an instability in a system started from a flat initial state
is qualitative because it is based on a  criterion that involves the {\em
average} value of the nearest-neighbor height difference $s_{max}$. Our
simulations show that the value of $s_{max}$ at a particular time $\tau$
exhibits large run-to-run variations. The distribution of $s_{max}$ 
shows a long tail extending to values substantially higher than the average
value. This variation in the value of $s_{max}$ is the main reason for the
large fluctuations in the calculated value of $\tau_{ins}$. 
Since an instability is expected to occur whenever the value of $s$
exceeds $h_c$, a system with a value of $\lambda$ which is lower than 
the $\lambda_c$ defined above in terms of the average value of $s_{max}$ would
exhibit the instability if a large value of $s$ lying in the tail of the 
distribution happens to be generated during the time evolution. Since the
tail of the distribution is more likely to be sampled if the system size is
large and if the simulation is continued for long times, the probability of
occurrence of an instability in a system with 
$\lambda < \lambda_c$ would increase with
system size and simulation time. These considerations show that a precise
definition of $\lambda_c$ is problematic. All we can say with certainty is that 
for sample sizes and simulation times used in typical numerical integrations
of growth equations
starting from a perfectly flat state, 
an instability would be very unlikely in a system exhibiting
conventional (rather than anomalous) 
dynamic scaling if the value of $\lambda$ is significantly
smaller than the critical value $\lambda_c$ defined above. As discussed above,
the results of our numerical investigation of the time of instability are
quite consistent with this prediction. 

We also note that the distinction we make between models
with $\lambda_c$ = 0 (e.g. the 1d LD equation) and those with $\lambda_c > 0$
(e.g. the KPZ equation and the 2d LD equation) 
is appropriate only when one considers the
evolution of the system from a flat initial state. Models with non-zero
values of $\lambda_c$ would show the instability (even if the value of
$\lambda$ is smaller than $\lambda_c$) if the initial state has a sufficiently
high pillar or a sufficiently deep groove (i.e. if $h_0 > h_c(\lambda)$, 
which is finite for any non-zero value of $\lambda)$. If the LD and KPZ 
equations are indeed the appropriate continuum descriptions of epitaxial
growth, as is currently believed \cite{rev1,rev2} to be the case, then our
finding of this generic instability for growth on a substrate with a high
pillar/groove may have important implications for real growth on patterned
substrates, which is a subject of considerable current interest in materials
science.

\section {Controlled Instability and Multiscaling}

In this section, we describe in detail  our numerical investigation 
of the connection between the instability described in the preceding
section and multiscaling behavior. The existence of such a
connection was suggested by the following observation made
in our numerical studies of
the discretized LD equation and the KD model in one dimension. We found
evidence for multiscaling (as indicated by the observed
time-dependence of the ratios $\sigma_q/\sigma_1$) during a short interval
of time immediately preceding the instability. During this time
interval, the interface width shows the expected scaling
behavior, but the functions $\{\sigma_q\}$ appear to scale with
different powers, the growth in time being faster for larger
values of $q$. This observation suggests that multiscaling
behavior may be closely related to this instability. We, therefore, carried
out a detailed investigation of this aspect.

It is easy to see that the instability described above would, in
general, lead to deviations from single-exponent scaling for the
quantities $\{ \sigma_q \}$. When the instability sets in, the value of
the nearest-neighbor height difference $s$ at the point of
instability becomes large and it grows rapidly in time. Since higher
moments of $s$ (i.e $\sigma_q$ for large $q$) are more 
sensitive to such large values of $s$, the growth of $\sigma_q$ in
time would be faster for larger values of $q$. The instability would also
produce a long tail extending to large values in the
distribution of $s$, leading to departures from single-exponent scaling for
the correlation functions $\{ G_q \}$. 
As mentioned above, we do find approximate multiscaling in our simulations
near the onset of the instability.
Typical data, obtained for the LD equation ($L$ = 10$^3$, $\lambda$ = 4.0,
100 runs) in one dimension, are shown in Fig.12. The growth of the ratio
$\sigma_q/\sigma_1$ with time is clearly seen, especially for large
values of $q$. We have shown data only upto $\tau$ = 100 because the 
instability is encountered at longer times and the time evolution of
the system can not be followed beyond the instability. 
Thus, the time interval over
which multiscaling is observed in the systems considered so far  
is very short. This is because the instability in
these systems is very ``strong'' in the following sense. 
In the discretized growth equations, the time evolution of the system can not 
be followed numerically beyond the instability time because the
height variables become too large. 
In the atomistic KD model, the height variables increase so
fast after the onset of the instability that global variables such as
the width of the interface begin to show deviations from scaling (see, for
example, Fig.3).
In order to explain the numerical
results obtained in Ref. \onlinecite{krug,dlkg,bdk}, it is
necessary to have a situation in which the global variables
scale in a normal way, whereas the quantities $\{\sigma_q\}$ and
$\{G_q\}$ show anomalous multiscaling. The discussion
above suggests that such a situation may be realized if the
instability is ``controlled'' in some way. We have considered
two different
classes of models with controlled instability. In models of the first
class, the instability is controlled by the introduction of 
terms with higher powers of the gradient of the 
height variable with appropriate coefficients.
The second class consists of atomistic models in which the
instability is controlled by modifying the deposition rule in a way that
restricts the development of large values of the nearest-neighbor
height difference. These models and the results obtained from numerical
studies of their behavior are described below. We emphasize that there is no 
unique way to control the growth instability, and, in principle, there must
be infinitely many different ways to do it. We have tried several simple
techniques of controlling the instability using minimal number of parameters,
as described in the next two subsections. In general, for a given atomistic
growth model (e.g. the DT model), the instability may be controlled in an
unknown and necessarily non-universal manner.

\subsection{Controlled Instability Models with Higher Powers of the Gradient}

The models we consider in this section are obtained by
replacing the $|\tilde{{\bf
\nabla}}h_i|^2$ term appearing in Eqs.(\ref{eqn5}) and (\ref{eqn11}) 
by $f(|\tilde{{\bf
\nabla}}h_i|^2)$, where $f$ is the following nonlinear function:
\begin{equation}
f(x) \equiv (1 - e^{-c x})/c. \label{eqn19}
\end{equation}
In the equation above, $c$ is an adjustable parameter.
Note that this replacement corresponds to the introduction of
an infinite number of higher-order nonlinear terms of the form
$|\tilde {{\bf \nabla}}h_i|^{2n}$ with specific coefficients
which depend on the value of $c$. 
In the following, we call this modified version of the discretized LD equation 
the {\em controlled Lai-Das Sarma} (CLD) model and the modified version of
the KD model is referred to as the {\em controlled Kim-Das Sarma} (CKD) model.
The function $f(x)$ behaves as $x$ for $x \ll 1/c$ and
approaches a constant value, $1/c$, in the limit $x \gg
1/c$. It is easy to show that the growth instability of
isolated pillars found in the original LD equation and the KD model 
is completely
suppressed if the value of $c$ is higher than a critical
value which depends on the value of $\lambda$. For values of
$c$ smaller than this critical value, the instability
occurs for an isolated pillar if its height lies within a  range 
$h_{min}(\lambda, c) < h_0 < h_{max}(\lambda, c)$.
This is shown in Fig.13 where we have plotted the quantity
\begin{equation}
\Delta(h_0) \equiv \frac{d}{d \tau} [h_n - h_{n-1}] \label{eqn20}
\end{equation}
for the noiseless CLD model for a state with a pillar 
of height $h_0$ at
the $n$-th site (i.e. for a configuration with $h_i = h_0$ for 
$i = n$ and $h_i = 0$ for all other $i$). The results shown are for $\lambda$
= 4.0 and three different values of the control parameter $c$. A positive
value of $\Delta$ implies that the pillar initially grows in time. As 
discussed in the preceding section, an initial growth of the height of a 
pillar is a necessary (but not sufficient) condition for the occurrence of
the instability. It is clear from the figure that there is no instability for
$c$ = 0.3, while the model would exhibit an instability for values of $h_0$
lying within a limited range for $c$ = 0.05 and 0.02. The range of $h_0$
values for which the instability is expected to occur clearly becomes
wider as the value of $c$ is decreased. Simulations carried out for the CLD
model with noise show an essentially similar behavior. The CKD model also
behaves in a very similar way. Fig.14 shows the probability of growth of an
isolated pillar of initial height $h_0$ in the CKD model with $\lambda$ = 2.0
and $c$ = 0.02. The growth probability is defined as before as the fraction of
runs in which the nearest-neighbor height difference at the site of the pillar
exceeds $h_0$. The growth probability is found to be close to zero at all
times for $h_0$ = 5 and $h_0$ = 45, indicating that these values are, 
respectively, lower than $h_{min}$ and higher than $h_{max}$. For $h_0$ =
25, on the other hand, the growth probability increases quickly to a value
close to unity and then falls off at long times. 
Thus, the instability for large value of $h_0$ is controlled 
in the new models by the introduction of an infinite series of higher powers
of the derivative with appropriate coefficients. This is
physically reasonable because terms involving higher powers
of the derivative are expected to come into the picture when the
derivative itself becomes large.

We have studied numerically the behavior of both CLD and CKD models for
different values of $\lambda$ and $c$. The numerical
divergence found in the original discretized LD equation disappears completely
for any non-zero value of $c$. This result confirms that
the instability we found in the original LD equation is a genuine
one, not an artifact of our numerics. The behavior we find in these two
models are qualitatively very similar. 
For  values of $c$ which are so large that  the
instability is completely absent, we find conventional scaling with
exponents close to the expected values. For very small values of
$c$, the instability is very ``strong'' and we find deviations
from scaling for global quantities such as the interface width.
Typical results are shown in Fig.15 where we have shown the time dependence
of the interface width $W$ for the CLD model (L = 10$^3$, $\lambda$ = 4.0)
with two values of $c$ (10 runs for $c$ = 0.02 and 40 runs for $c$ = 0.05), and
for the CKD model ($L$ = 10$^3$, $\lambda$ = 2.0), also with two different
values of $c$ ($c$ = 0.02, 200 runs and $c$ = 0.005, 10 runs). In both
models, the results for the larger value of $c$ show the expected power-law
scaling in time, whereas the data for the smaller value of $c$ exhibit a
``strong'' instability at which the scaling behavior of $W$ breaks down.
It is interesting to note that the growth of $W$ before the occurrence of
the break in the curves for the smaller values of $c$ is almost
indistinguishable from that seen for the larger values of $c$ for 
which the power-law growth continues till long times. 
Thus, the scaling behavior of $W$ even in the ``strongly'' unstable
situation mimics the ordinary power-law growth upto the instability onset 
time which may be very long, depending on the values of the parameters in
the model. As noted in section III above, similar results were obtained for
the original (uncontrolled) models also. The plots in Fig.15 
show explicitly the similarity between the behavior of the two models.
We describe below results obtained
for the atomistic CKD model because simulations
of this model are easier, so that better statistics can be
obtained. Very similar results, but with poorer statistics, were
obtained for the CLD model.

Interesting behavior is found in simulations of the CKD model with
intermediate values of $c$ for which the instability occurs
for a limited range of values of $h_0$. For such values of
$c$, the instability is expected to be operative over a
limited time interval. At very early times, the values of $s$
are small and no instability occurs. As time progresses, the
instability sets in when the value of $s_{max}$ crosses
$h_{min}$. 
The value of $s$ at the point of instability
grows rapidly in time until the growth is cutoff at $h_{max}$. 
At subsequent times, the instability occurs at more
and more points in the system. 
The number of points at which a
fresh instability can occur decreases in this process. Also, effects of
the instability become less pronounced at long times because the typical
value of $s$, which increases with time even if there is no instability,
becomes comparable to $h_{max}$ at
sufficiently long times. So, the
instability is expected to become ineffective at long times. If
multiscaling arises due to the instability, then one expects to
see multiscaling only during the finite time interval over which
the instability is active. This is precisely the behavior we find
in the simulations. In Figs.16 and 17, we show a representative
set of simulation results obtained for $L$ = 1000, $\lambda$ =
2.0, and $c$ = 0.02, averaged over 2000 runs. For these values of $\lambda$
and $c$, $h_{min} \simeq $ 6.0 and $h_{max} \simeq$ 34.0. As shown in
Fig.16, the rms interface width $W$ shows excellent scaling with
an exponent close to 1/3. The quantities $\{\sigma_q\}$, however,
show clear evidence of multiscaling during the time interval
between $\tau \approx$ 5 and $\tau \approx$ 1000.
Power-law fits to the data over this time interval yield
the following values for the effective exponents: $\alpha_1/z$ =
0.14 $\pm$ 0.02, $\alpha_2/z$ = 0.17 $\pm$ 0.02, $\alpha_3/z$ = 0.22 $\pm$
0.02, $\alpha_4/z$ = 0.26 $\pm$ $0.03$. These exponent values are
similar to those found in Ref. \onlinecite{krug} for the 
1d DT model. As shown in
the inset of Fig.16 where we have plotted the time-dependence of
the ratios $\sigma_q/\sigma_1$ for $q$ =  2, 3 and 4, the multiscaling is
not present at very early times and also at times longer than about
1000. By monitoring the time development of $s_{max}$, we find
that $\tau \approx $ 1000 is precisely the time at which the 
instability begins to level off. We have carried out much longer runs for
smaller samples and found that multiscaling of $\{\sigma_q\}$ is absent 
for all $\tau >$ 1000. 

In Fig.17, we have plotted the correlation functions $\{ G_q \}$
for the same system at time $\tau$ = 1000. Multiscaling is
clearly seen, with the following exponent values calculated from
power-law fits to the data for $2 \le l \le 10$: $\zeta_1$ = 0.74 $\pm$
0.03, $\zeta_2$ = 0.66 $\pm$ 0.03, $\zeta_3$ = 0.58 $\pm$
0.03, and $\zeta_4$ = 0.50 $\pm$ 0.03. These exponent values are also
similar to those found in Ref. \onlinecite{krug} for the 1d DT model. 
The multiscaling
behavior for $l \le$ 20 is clearly illustrated in the inset where
we have plotted the ratios $G_q(l)/G_1(l)$ for $q$ = 2, 3 and 4
as functions of $l$. 

We have also investigated the dependence of $W$ on the sample-size $L$
in this model for small values of $L$. The time-dependence of $W$ in a
sample of length $L$ is expected to be described by the finite-size scaling 
equation
\begin{equation}
W(\tau, L) =  \tau^{\beta} f(L \tau^{-1/z}).
\label{eqn21}
\end{equation}
The scaling function $f(x)$ in Eq.(\ref{eqn21}) goes to a constant as
$x \rightarrow \infty \,\, (\tau \ll L^z)$, so that the interface width
grows as $\tau^{\beta}$ at early times. From the data shown in Fig.16, we
estimate the value of $\beta$ to be close to 0.35. The scaling function
behaves as $f(x) \approx x^{\beta z}$ in the $x \rightarrow 0$ limit, so
that at times long compared to $L^z$, the interface width becomes time-
independent and proportional to $L^{\alpha}$ with $\alpha = \beta z$. The
value of the exponent $\alpha$ is estimated to be close to 1.25 form the
observed sample-size dependence of the interface width at saturation. These
exponent values correspond to $z \simeq$ 3.5. The scaling equation 
(\ref{eqn21}) can also be written as
\begin{equation}
W(\tau, L) L^{\alpha} = g(\tau L^{-\alpha/\beta})
\label{eqn22}
\end{equation}
so that for a proper choice of the values of the exponents $\alpha$ and
$\beta$, the data for different $L$ and $\tau$ should collapse to a single
scaling curve when $W L^{\alpha}$ is plotted against $\tau L^{-\alpha/\beta}$.
A scaling collapse of the data obtained for three different values (20, 40
and 80) of $L$ for the CKD model with $\lambda$ = 2.0 and $c$ = 0.02 is shown
in Fig.18. The number of independent runs used in generating the data shown
is 2000 for $L$ = 20 and 1000 for $L$ =40 and 80. The exponent values used in
the scaling plot are $\alpha$ = 1.25 and $\beta$ = 0.355. Thus, it is clear
that the CKD model with these parameters exhibits the expected scaling 
behavior for the global quantity $W$.  
These exponent values, when combined with the values of the 
exponents $\alpha_q/z$ and $\zeta_q$ quoted above, satisfy the expected
relation $\alpha_q + \zeta_q = \alpha$ within error bars,
although there appear to be systematic deviations from this
relation for large $q$. Very similar results were obtained in
Ref. \onlinecite{krug} for the DT model.

We have also calculated the distribution of the nearest-neighbor height 
difference $s$ at long times. Results for the CKD model with $\lambda$ = 2.0 
and $c$ = 0.02 are shown in Fig.19. We show the distribution $P(s)$ for 
two different cases: for a system with $L$ = 10$^3$
(averaged over 2000 runs) at time $\tau$ = 1000 (the time at which 
multiscaling in $\{\sigma_q\}$ 
tapers off, see Fig.16), and for a system with $L$ = 80
(1000 runs) averaging over the time interval $5 \times 10^4 \le \tau \le
5 \times 10^5$ in the saturation regime. In both cases, the distribution
of $s$ is found to be strongly non-Gaussian with a long tail extending to large
values of $s$. As shown in the figure, a power-law form, $P(s) \propto 
s^{-\eta}$, provides an excellent fit to the data over more than four
decades. The best-fit 
value of the exponent $\eta$ of the power law is found to be
3.2 for the $L$ = 1000 system and 2.5 for the $L$ = 80 system. This power-law
behavior is different from the result obtained \cite{krug} for the 1d DT
model in which $P(s)$ appears to show 
a stretched exponential behavior. As shown in
the inset of Fig.19, a stretched exponential form with a stretching exponent
$\approx$ 0.6 provides a good fit to our $L$ = 80 data 
for small values of $s$, but
fails at larger values.

These results clearly show that multiscaling behavior very similar to
that observed in Ref \onlinecite{krug,dlkg,bdk} can be generated
by a controlled instability of the kind described above. It
should, however, be noted that the multiscaling we find is
transient in the sense that it occurs only over a limited
range of time. A careful look at the data of Ref
\onlinecite{krug,dlkg,bdk} suggests that the same is true
for the atomistic models studied in these papers. The
approximate multiscaling we find is non-universal: the
effective exponents $\zeta_q$ and $\alpha_q$ extracted from our
numerical data seem to depend on the way in which the
instability is controlled. Similar non-universality is also
found in the atomistic models studied in Ref \onlinecite{krug,dlkg,bdk}.

The similarity between the CKD model with appropriate choice of the
parameters and the DT model is also illustrated by a comparison between the
growth profiles in the two models. Since both these models are atomistic in
nature, it makes sense to compare the growth profiles obtained at the same
value of the discrete time measured in units of number of layers deposited. 
In Fig.20 and Fig.21, we show typical growth profiles in, respectively,
the 1d DT model and the
1d CKD model with $\lambda$ = 2.0 and $c$ = 0.02. Both profiles are obtained
for samples with $L$ = 1000 after the deposition of 10$^4$ layers. The
average value has been subtracted off from the heights
plotted in these Figures. The
similarity between these two profiles becomes evident when one is inverted
relative to the other i.e. if the transformation $h \rightarrow -h$ is made
in one of the profiles. This transformation is equivalent to changing the
sign of $\lambda$ in the CKD model. Thus, the DT model appears to be similar
to the CKD model (and also to
the CLD model) with a negative value of $\lambda$. The
asymmetry between the peaks and troughs of the profile is evident in both 
Fig.20 and Fig.21. In the 1d DT model (Fig.20), the peaks of the profile are
generally 
rounded and the troughs tend to be very sharp and ``spiky''. The profile also 
wanders a longer distance from the baseline (average height) in the 
negative direction. Both these features are reversed in the profile  obtained
in the 1d CKD model (Fig.21). The reason for this ``inversion'' is quite
simple. The profile obtained in the 
CKD model with positive $\lambda$ exhibits sharp peaks and wanders a longer
distance on the positive side because pillars are unstable in this model.
The situation is reversed in the DT model because, as shown below, grooves
have a finite probability of getting deeper in this model. This difference
can be eliminated simply by changing the sign of $\lambda$ in the CKD model. 

\subsection{Controlled Instability Models with Modified Deposition Rule}
 
We have constructed and studied by simulations a modified version of the
atomistic KD model in which the deposition rule is changed in order to
control the growth of nearest-neighbor height differences. This study
was motivated by the following observation. In order to explore further the
connection between controlled instability and multiscaling, we studied
by simulations the evolution of isolated
pillars and grooves in the 1d DT model. We start with
a configuration which is flat everywhere except at the central point where 
there is a pillar of height $h_0$ or a groove of depth $h_0$. We then
simulate the time evolution of this state and measure the probability that
the absolute value of the nearest-neighbor height difference at the central
site exceeds $h_0$ at time $\tau$. We find that the probability of a pillar
becoming higher is strictly zero whereas grooves have a non-zero probability
of becoming deeper. Thus, the asymmetry between
grooves and pillars found in the models described above is present in the 
DT model also. It is not difficult to explain the origin of this asymmetry.
Consider a configuration in which $h_i$ is zero everywhere except at the 
site $n$ where $h_n = h_0$. For a positive $h_0$ (a pillar at site $n$), a
particle deposited at site $n$ diffuses to one of its nearest neighbor sites
because the number of bonds at site $n$ is one whereas the number of bonds
at the sites $n+1$ and $n-1$ is two. Particles deposited at sites $n+1$ and
$n-1$ do not diffuse because each of 
these two sites have two bonds. Thus, there is no
deposition sequence which can increase the height differences $|h_n - h_{n-1}|$
and $|h_n - h_{n+1}|$. 
In contrast, for a negative $h_0$ (a groove at site $n$), there
are certain deposition sequences which increase these height differences. 
Consider, for example, the sequence in which a particle is first deposited 
at the site $n+2$ and then another particle is deposited at the site $n+1$.
The particle deposited at $n+2$ stays there because although this site
has only one bond, the nearest neighbors of this site also have only one
bond each. The particle deposited subsequently at $n+1$ also stays at this
site because it now has two bonds. 
The difference between the heights at
sites $n$ and $n+1$ increases in this process. Therefore, the probability of
a groove becoming deeper in the course of time should be non-zero in this model.
This simple picture also 
implies that this probability  
should not depend strongly on $h_0$, the initial depth of the groove, as long
as $h_0$ is not very small. We believe that the non-zero probability of
grooves becoming deeper is the basic reason for the occurrence of large 
values of the nearest-neighbor height difference (which lead to multiscaling)
in the 1d DT model. This belief is supported by examinations of the 
height profiles generated
in simulations of the 1d DT model which show that large values of
the nearest-neighbor height difference almost always 
correspond to deep grooves in this system (see, for example, Fig.20).

The results of our simulations on the 1d DT model, obtained by averaging over
10$^5$ runs on a $L$ = 32 sample, 
are shown in Fig.22. The data shown were obtained for $h_0$ = 100. The 
same behavior is found for
all values of $h_0 > 10$. 
These results are consistent with the simple picture described above.
The probability of increase of the depth
of a groove is found to increase initially with time, reach a maximum near
$\tau = 1$ and decay slowly to zero at longer times.
The shape of the 
probability vs. time curves in this model is qualitatively similar to that
of the curve shown in Fig.14 for $h_0$ = 25 in the CKD model. There are,
however, large quantitative differences between the two curves. 
Clearly, the probability of growth of the nearest-neighbor 
height difference is substantially smaller
and decays faster in time in the DT model. 
We have also investigated the time
evolution of wider grooves in the 1d DT model. A groove of depth $h_0$ and 
width $w$ corresponds to an initial configuration in which the height is
$- h_0$ at the sites $n, n+1,\cdots, n+w-1$, and zero everywhere
else. All values of $n$ are equivalent because 
periodic boundary conditions are used. 
From simulations of the time evolution of initial configurations with
different values of $w$ and $h_0$, we calculate the probability that the
difference between the height at the central site of the groove and the 
average height outside the groove is greater than $h_0$ at time $\tau$. 
As shown in the inset of Fig.22, where we have plotted the results 
for $w$ = 3 and $h_0 = $ 30 and 90 (these results were 
obtained by averaging over
1000 runs on samples with $L$ = 128), this probability is quite 
high and it decays rather 
slowly with time. The probability of a groove getting deeper is found to 
increase with increasing $w$ and $h_0$. The decay of this probability in
time becomes slower as the values of $w$ and/or $h_0$ are increased. Detailed
examination of the configurations generated in the simulations shows that the
difference between the height at the center of the groove and the average
height outside the groove does not increase much beyond $h_0$, but remains
slightly higher than $h_0$ with a high probability over a relatively long
period of time which increases with $w$ and $h_0$. This behavior may be 
understood in the following way. Consider a groove with $w$ = 3 centered at
the site $n$. Each of the nearest neighbors of the central site
$n$ has two bonds in the
initial state. Therefore, particles
dropped at one of these two neighbors of the central site stay at
that site, and the central site does not get any particle from its nearest
neighbors during the initial time evolution of the system. There is, however,
a finite probability for a particle dropped at the central site to move to
one of the neighboring sites. This would happen, for example, if the heights
at the sites $n$, $n+1$ and $n-1$ are equal. So, the rate at which the height
at the central site grows initially is slightly lower than the rate of
growth of the average height (which, by definition, is one layer per unit time
in this model). These considerations do not apply at long times when the 
groove begins to fill up. For this reason, the probability of the groove 
getting deeper begins to decrease at long times, making this ``unstable''
behavior a long-lasting transient. The slow decay of this transient behavior
for large values of $w$ and $h_0$
may have important implications for real growth on patterned substrates.

An important difference
between the results obtained for the DT and CKD models 
is that the growth probability
is close to zero for $h_0 > h_{max}$ in the CKD model, whereas it is 
almost independent of $h_0$ (for $w$ = 1) or an increasing function of
$h_0$ (for $w \ge 3$) in the DT model.
This is probably the reason why the 1d DT model exhibits multiscaling over
a longer period of time than the CKD model (in the 1d DT model, multiscaling
lasts for at least six decades in time \cite{krug,dlkg,bdk}, compared to about 
three decades in the CKD model). 
This observation suggests that 
a controlled instability model in which the growth probability remains non-zero
for large values of $h_0$ may exhibit multiscaling behavior over a longer 
period of time. It is difficult to construct such a model
along the lines described in section IVA. This is because an
increase in the value of $h_{max}$, which can be achieved 
either by increasing $\lambda$ or by
decreasing $c$, makes the instability strong, leading to deviations from
scaling for global quantities such as $W$ (see, for example, Fig.15). 
We, therefore, constructed a 
different class of atomistic models in which the
instability in the original KD model 
is controlled by an appropriate modification of the deposition rules.
We describe below one such model and the results obtained from simulations
of this model in one dimension.

The deposition rule in this model is designed to restrict the development
of large values of the nearest-neighbor height difference. As in the KD
model, the height variables in this model are discrete and time is measured
in units of number of layers deposited. The deposition of a particle involves
the following steps. A site is chosen at random and the KD rules (described
in section II) are used to
determine whether a particle is to be added to the chosen site or to one of
its nearest neighbors. Let $n$ be the index of the site which is selected 
for the addition of a particle according to the KD rules 
and let $s_l \equiv |h_n - h_{n-1}|$ 
and $s_r \equiv |h_{n+1} - h_n|$ be the 
nearest-neighbor height differences
at this site before the addition of the particle. The addition of a particle
at site $n$ would change the values of both $s_l$ and $s_r$.
If both $s_l$ and $s_r$ would decrease due to the addition of the
particle at site $n$, the particle is added 
at site $n$ with probability one. If one of
these two height differences would increase 
while the other one would decrease by the
addition of a particle at site $n$, then we define $S$ to be the
value of the height difference which would increase. If both the height
differences increase due to the addition of a particle at site $n$, then
we define $S$ to be the larger of $s_l$ and $s_r$.
In these cases, the particle is deposited at site $n$ with probability
$p = \exp(-u S)$, where $u$ is an adjustable parameter. Operationally,
this is done by generating a random number $r$ which is distributed 
uniformly between 0 and 1 and the particle is deposited at site $n$ (i.e.
the height variable at site $n$ is incremented by one) if $r \le p$. If $r 
> p$, one of the nearest-neighbors of the site $n$ are chosen randomly and
the particle is deposited there. This model reduces to the original KD model
for $u$ = 0. For small positive values of $u$, the modification in the 
deposition rule disfavors but does not completely eliminate the growth of 
large nearest-neighbor height differences. There is a second route by which
large height differences may form in this model. When a nearest-neighbor
of the site $n$ is chosen for the deposition of a particle by the stochastic
rule described above, it is not checked whether this deposition would increase
the height differences between this site and its nearest neighbors. So, in
some cases, the deposition of a particle at one of the nearest-neighbors of
the site $n$ leads to an increase in the value of some nearest-neighbor 
height difference, and such increases are not controlled by the parameter $u$.
One can, in principle, include such checks in the deposition rule, but this
would make the rule very complicated and difficult to simulate efficiently. 

Fig.23 shows our simulation results for the probability of growth of an
isolated pillar of initial height $h_0$ in this model with $\lambda$ = 2.0
and $u$ = 0.06. The data shown were obtained by averaging over 5000 runs for
systems with $L$ = 100. As expected, the probability of growth of
isolated pillars in this model decreases with increasing $h_0$, but does not
go to zero for large values of $h_0$. The magnitude of the 
growth probability for large values
of $h_0$ in this model is similar to that in the 1d DT model, but the 
probability decays faster in time in the DT model. The observation of a 
non-zero probability of growth of pillars with large values of $h_0$ suggests
that the time interval over which this model exhibits multiscaling behavior 
should be longer than that for the CKD model with similar parameter values. 
This expectation is confirmed by our simulation results which are shown in
Figs.24 and 25 for a system with $L$ = 1000, $\lambda$ = 2.0 and $u$ = 0.06. The
data shown represent averages over 200 runs. As can be seen in Fig.24, the 
time-dependence of the rms interface width $W$ shows excellent power-law scaling
with $\beta \simeq$ 0.35. The quantities $\{\sigma_q\}$, on the other hand,
show evidence of multiexponent scaling over the entire time interval of
these simulations. This time interval (10$^4$ units) is an order
of magnitude longer than the time interval over which multiscaling was observed
in the CKD model (see Fig.16). The fact that the $\{\sigma_q\}$ with different
values of $q$ do not grow in time with the same exponent is clearly shown
in the inset of
Fig.24 where we have plotted the time-dependence of the ratios $\sigma_q/
\sigma_1$ for $q$ = 2, 3 and 4. The observed increase of the values of these  
ratios with time implies that $\{\sigma_q\}$ grows faster in time for larger
values of $q$. It is perhaps more appropriate to characterize the observed
behavior of the $\{\sigma_q\}$ as {\it deviation from single-exponent scaling},
rather than multiscaling. This is because log-log plots of $\sigma_q$ vs
$\tau$ (Fig.24) show substantial deviations from linear behavior. For this
reason, it would not be particularly
meaningful to define the exponents $\alpha_q/z$ in
this model. As noted above, similar (but less pronounced) 
deviations from pure power-law
behavior of the $\sigma_q$s 
have also been observed in simulations of the 1d DT model. The data for the
correlation functions $\{G_q(l)\}$ measured at time $\tau$ = 10$^4$ are
shown in Fig.25. These data
show clear evidence of multiscaling, with exponent values $\zeta_1 \simeq
$0.79, $\zeta_2 \simeq$ 0.71, $\zeta_3 \simeq$ 0.62, and $\zeta_4 \simeq$
0.57. These results show that multiscaling similar to that observed in the
1d DT model can be generated by controlling the instability in the 1d KD 
model by appropriate modifications of the deposition rules.

A typical growth profile obtained in the modified KD model with 
$L$ = 1000, $\lambda$ =
2.0 and $u$ = 0.06 after the deposition of 10$^4$ layers is shown in Fig.26.
The similarity between this profile and the one shown in Fig.20 for the 1d
DT model (after making the $h \rightarrow -h$ transformation) is quite
remarkable. All the characteristic features present in the DT model profile are
also present in the profile obtained in the modified KD model. This similarity
provides additional support to the connection we propose between the  DT
model and the models with controlled instability studied in this paper. One
could perhaps attempt to obtain better quantitative agreement between the
multiscaling behavior in the DT and the controlled instability models by
trying other variants of the LD equation and the KD model (e.g. by trying other
forms for the function $f(x)$ in Eq.(\ref{eqn19}) for the CLD/CKD model or by
modifying the deposition rules of the KD model in a different way). 
However, we do not
see much purpose in such an attempt because (i) the multiscaling behavior is
non-universal as we have been emphasizing throughout this paper, and (ii) such
efforts are bound to be computationally intensive, the number of possible
modifications of the LD equation and/or the KD model being arbitrarily large. We
believe that the results presented in this paper unambiguously establish the
(qualitative) connection we propose between controlled instability and
the observed multiscaling in the 1d DT model.

\section{Discussions}

In this section, we discuss a number of questions that are raised by the 
results obtained in our study.
The work described here leads to two main results - the first one is about the
presence of an 
instability in discretized growth equations and the second one is about
the connection between this instability and multiscaling. Both these results
have important implications in the study of models of surface growth. The
observation of an instability in discretized versions of nonlinear 
growth equations 
raises the question of whether a similar instability is also present in the 
truly continuum limit. As discussed in section III above, the presence of
an instability in the discretized version of the 1d noiseless KPZ equation
{\em does not} reflect the behavior of the corresponding 
continuum KPZ equation which does not
have any instability. It is not clear 
whether the same conclusion would apply to the other growth equations
(the KPZ equation with noise and the LD equation) considered here, mainly due
to the fact that no exact result is available about the behavior of these
continuum equations.  The work of Ref. \onlinecite{pbk} suggests that 
the noiseless
continuum LD equation in one dimension does not exhibit a true finite-time
singularity in the sense that the height variable remains bounded by a value
proportional to a power of the system size. That work, however, does
not rule out the occurrence of an instability in which the height variable
increases very rapidly in time at one or more isolated points in the system.
In fact, the numerical results reported in Ref. \onlinecite{pbk}
indicate that pillar-like structures initially grow in height in the continuum
system also. 
Throughout this paper, 
we use the term ``instability'' to mean a rapid growth of the 
magnitude of the height variable
in a local region of the system. The rate of change of the height in the region
of the instability must be  much faster than the corresponding rate in the
background. The rapid change of the height in the instability region need not
lead to a true divergence. The question of whether a true divergence occurs or
not is interesting, but 
not very important for most practical purposes. For example, a rapid
growth of the height variable at one or more points in the sample may
lead to deviations from scaling behavior
even if there is no true divergence. This is shown clearly in Fig.15 above.
Also, it is almost impossible to 
distinguish in numerical work 
between a true divergence and a growth to a very large but finite value. The
issue of a true divergence in the continuum growth equations is beyond the 
scope of our work. 

Several atomistic models which are believed to belong to 
the universality class of the KPZ equation or the LD equation are known 
not to have any instability. For example, the restricted solid-on-solid model
of Kim and Kosterlitz \cite{kk}, 
which is believed to be in the same universality
class as the KPZ equation, can not exhibit any instability of the kind being
considered here because the nearest-neighbor height difference can not,
by construction, exceed
unity in this model. Similarly, the recently introduced 
conserved version \cite{kkk} of the 
Kim-Kosterlitz model, which is supposed to belong to the universality class
of the LD equation, also does not exhibit any 
instability. Another example is the
model solved exactly by Gwa and Spohn \cite{gs} in one dimension.
This model, believed to belong to the 1d KPZ
universality class, is also known not to 
have any instability. These results, however, {\em do not} 
provide much help regarding
the possibility of the occurrence of an instability in the corresponding
continuum growth equations. This is because
these models certainly differ from the corresponding continuum growth
equations by virtue of 
the presence of terms which are irrelevant in the renormalization
group sense. As shown in section IV above (and discussed in more detail
below), the presence of such irrelevant
terms may control or eliminate altogether the instability found in 
discretized growth equations. Therefore, the absence of an instability
in models which are  in the same universality class as the continuum growth
equations {\em does not} imply that the growth equations do not have any
instability.

It is clear from the discussion above that the possibility of occurrence of an
instability in the continuum growth equations remains an open problem. We hope
that our work will stimulate further investigation of this question. This
question acquires special importance in view of the fact that
a large part of the currently available information about the behavior 
of nonlinear growth equations has been obtained from studies of discretized 
versions. Our work, especially the results obtained here (and also in Ref.
\onlinecite{nb}) about the difference between the behavior of the discrete
and continuum versions of the noiseless 1d KPZ equation, 
raise serious questions 
about the applicability of the information obtained from numerical integration
of discretized growth equations to the continuum equations. Our simulations
suggest that the results obtained from numerical integration of the discretized 
equations should apply to the continuum case {\em as long as there is no
instability}. This conclusion is suggested by the observation 
that the values of the ``global'' exponents $\beta, z$
and $\alpha$ extracted from the results of
direct integration of discretized growth equations before the
occurrence of the instability (or from runs in which no instability
occurs, either due to the smallness of $\lambda$ or due to the imposition of
control) are in good agreement (within error bars) with the exact (for
the 1d KPZ equation) or expected (from e.g. renormalization group calculations
for the LD equation) results for the continuum equation. We quote here the
values of the exponent $\beta$ which is determined most accurately in our
work. The calculated value of $\beta$ for the discretized LD equation in 
one dimension is 0.325 $\pm$ 0.01, and that for the discretized 1d KPZ
equation is 0.32 $\pm$ 0.02. 
These values are in good agreement with the expected result,
$\beta$ = 1/3, for the 1d LD equation and the exact result, $\beta$ = 1/3, for
the 1d KPZ equation. It would be useful to
substantiate this conclusion with further study.

Our work also brings out the importance of terms which are often
not included in continuum 
growth equations because they are irrelevant in the 
renormalization group sense. These terms involve higher powers of the gradient
of the height variable. It is well-known that these terms are irrelevant in the
KPZ equation in one dimension 
and in the LD equation in two and higher dimensions.
This, however, does
not mean that they are totally unimportant. The results described in 
section IV
about the possibility
of controlling the 
instability in the discretized growth equations by the
introduction of higher powers of the gradient with appropriate coefficients
indicate that
such terms may play a very important role in the stability of the growth
equations. These terms do not affect the values of the
global exponents, but the growth equation may be unstable if such terms are not
included. A recent paper by Marsili and Bray \cite{mb} makes a
similar point. They consider an infinite-range version of the KPZ equation
and show that this equation is unstable if certain higher-order terms 
(which are irrelevant in the RG sense) are not included. The instability
they find is similar to the one we have found in discretized growth equations.
There are well-known examples in equilibrium critical phenomena where formally
irrelevant terms (the so-called ``dangerous irrelevant variables'')
play a similar role in the control of 
instabilities. Consider, for example, the standard Ginzburg-Landau free energy
functional \cite{ma}
for a scalar order parameter (the so-called $\phi^4$ field theory).
It is well-known \cite{ma} that the $\phi^4$ term in this free energy functional
is irrelevant in the renormalization group sense if the dimension of space is
higher than 4.
However, the {\em sign} of the coefficient $u$ of the $\phi^4$ term is very
important because a negative value of $u$ leads to an instability. To control
this instability, one needs to introduce one or more
higher-order terms (typically, a $\phi^6$
term) which are also irrelevant in the renormalization group sense. The
role played by the $\phi^6$ term in this example is qualitatively similar to 
the role of the higher-order terms introduced in our models of controlled
instability. Another aspect of this example \cite{ma}
from equilibrium critical phenomena
may also be relevant to the growth problems being considered here.
The Ginzburg-Landau model is known to exhibit strong crossover effects
if it is close to an instability i.e. if the
value of $u$ is positive but very small, the free energy density being
singular in the limit $u \rightarrow 0$. 
This crossover in this model 
is from apparent
tricritical behavior (which would be the asymptotic critical behavior for $u$
= 0) to the usual critical behavior expected for $u > 0$. These crossover
effects lead to apparent non-universal values of the effective critical
exponents, and the correct values of the critical exponents are observed only
very close to the critical point (i.e. at very long length scales). As
discussed below, there are reasons to believe that the non-universal 
behavior observed in some of the growth models studied in our work 
is also a consequence of crossover effects arising from the proximity of
the system to an instability. It will be particularly interesting to try
to establish formally this 
suggestive but qualitative analogy between anomalous {\em dynamic} scaling
in DT and controlled instability models and crossover scaling induced by 
dangerous irrelevant variables
in equilibrium critical phenomena.

The observation that the instability in discretized growth equations can be
effectively controlled or eliminated altogether 
by the introduction of formally irrelevant higher-order
terms suggests a practical solution to the numerical problems encountered
in previous studies \cite{tu,mkw} using direct integration. If the
discretized version of a continuum growth equation belongs in the same 
universality class as the continuum equation itself, then
a direct numerical integration of a version of the discrete
equation which is controlled by the introduction of irrelevant higher order 
terms should yield values of global critical exponents appropriate for the
continuum growth equation without running into instability problems. The
usefulness of this method is illustrated by the results described in 
section IVA for models in which the instability is controlled or eliminated.  
It is interesting to note that the prescription suggested by Newman and
Bray \cite{nb} for getting rid of the instability in the discretized version
of the noiseless 1d KPZ equation, while differing in details,
also amounts to the introduction of terms
involving higher powers of the gradient of the height variable.

The 1d LD equation is special in the sense that the terms involving higher 
powers of the gradient are all {\it marginally relevant} 
in one dimension. It has been
argued in Ref. \onlinecite{bdk} that the marginal relevance of these terms may 
lead to nonuniversal corrections to the critical 
exponents and also to anomalous dynamic scaling where local and global
exponents differ \cite{dgk}.
The value of the exponent $\beta$ found in our simulations of the CKD model
and the modified KD model described in section IVB is $0.355 \pm 0.005$, which
is close to, but significantly different from the value 1/3 expected from 
renormalization group calculations \cite{dsk} on the LD equation. It is
tempting to attribute this difference to the presence of higher powers of the
gradient in the controlled KD models. However, we do not have 
any other evidence 
to back up this explanation which, therefore, remains speculative.
The leading order mode-coupling analysis
of Ref. \onlinecite{bdk} also suggests that the marginal relevance
of the 
higher order terms may play an important role in the multiscaling 
behavior. As described in section IVA, we do find approximate multiscaling in
models in which the instability is controlled by the introduction of an 
infinite series of higher powers of the gradient with appropriate coefficients.
Therefore, there is a definite connection between multiscaling and the
coefficients of the higher order terms. 
However, the values of the effective multiscaling exponents $\alpha_q$ and
$\zeta_q$ obtained from our simulations 
do not satisfy the specific quantitative prediction of Ref. \onlinecite{bdk}.
Thus, on a conceptual level, there is a correspondence between our finding of
a (``controlled'') instability and the infrared singularity underlying the 
work of Ref.\cite{bdk}. However, a more detailed connection between these two
works must await further investigation.

The possibility that continuum and discretized versions of a growth
equation may exhibit different behavior brings up the question of which
version is more appropriate for describing real physical systems.
While we do not claim to have an answer to this question, we wish to point
out that it may be inappropriate to regard the discretized version of a growth
equation to be less ``fundamental'' than the continuum version.
Discretized versions of nonlinear growth equations may actually be 
closer to the physics of growth processes than the continuum equations. 
This is because all growth processes are discrete at the atomic scale 
due to the presence of the cutoff introduced by the atomic lattice structure.
A continuum description is obtained under certain assumptions about the
smoothness of the growth profile. The discreteness at the atomic
level is incorporated in the continuum description through the
introduction of a short-distance cutoff. The instability we find suggests
that the assumptions which go into the development of a continuum
description may not be valid under certain circumstances, depending
on the values of the bare coupling constants, cutoffs etc.

Finally, we discuss the implications of our results on the origin and nature
of multiscaling in models of surface growth. It is clear from the
results described in section
IV that a controlled instability is responsible for 
the multiscaling behavior in the models we have studied.
In the CKD model described in section IVA,
multiscaling is found only if the value of the control
parameter $c$ is such that the instability is present. Also, multiscaling 
behavior for such values of $c$ is observed only during the time
interval over which the instability is operative. The same behavior is found 
for the modified KD model described in section IVB.
The multiscaling behavior exhibited by these models
is very similar to that observed 
\cite{krug,dlkg,bdk} in the 1d DT model and in
other related atomistic models of surface growth.
As described in section IVB,
the behavior of the probability of growth 
of an isolated groove as a function of time in the 1d DT model is 
similar to that found in the modified KD model for appropriate values of the
control parameter $u$. 
All these results 
clearly establish a connection between multiscaling in growth models
and a controlled instability of the kind described here. Our study also leads
to the important conclusion 
that the multiscaling found in growth models is necessarily 
{\em non-universal} and {\em
transient} in time. Consider, for example, the CKD model described in section
IVA. 
In this model,
the values of the exponents $\beta$, $z$ and $\alpha$, which 
describe the {\em global} properties of the growing interface, are essentially
the same 
for all values of the control parameter $c$ (of course, the value of $c$
should be sufficiently large so that
the rms surface width $W$ does not show any departure from power-law 
behavior; otherwise, these exponents can not be defined). 
The models with different values of $c$, therefore, belong to the
same universality class as far as the global behavior is concerned. On the
other hand, the
multiscaling behavior, which involves {\em local} quantities because it is
manifested in the time-dependence of the moments of the nearest-neighbor
height difference and in the dependence of the correlation functions $G_q(l,
\tau)$ on $l$
for $l < \xi(\tau)$, is found to be very different for different values of $c$.
Similar results are obtained for the modified KD model of section IVB. In this
model also, 
the global exponents are found to be insensitive to the value of the control
parameter $u$, but the multiscaling behavior depends crucially on it.  
These results clearly show that multiscaling in these models is a non-universal
feature. We also find that all the models studied in our work exhibit
multiscaling only over a limited period of time. The length of the time
interval over which multiscaling is observed varies greatly from one
model to another, but multiscaling is found to disappear at sufficiently
long times in all these models. This is {\em not} a saturation 
effect because the 
rms surface width $W$ continues to grow beyond the time at which
multiscaling disappears (see Fig.16, for example). A careful look at the
simulation data of Ref.\cite{krug,dlkg,bdk} shows that the feature of
non-universality and the transient nature of multiscaling are present
in varying degrees in all the atomistic models studied in these papers. It is,
therefore, reasonable to conclude that multiscaling in discrete
growth models is
a non-universal transient behavior, possibly related to the presence of 
higher-order terms involving higher powers of the gradient of the height
variable in the continuum equations appropriate for these discrete models.
The similarity between our results for the controlled-instability versions  
of the KD model (which, by construction \cite{kds}, provides
an atomistic version of the
LD growth equation) and those obtained in Ref.\cite{krug,dlkg,bdk} for the
1d DT and related models suggests that the latter models are described by the
LD equation with the addition of terms containing higher powers of the 
height gradient
with appropriate coefficients. The marginality of these terms in one dimension
may provide an explanation of why the time period over which multiscaling is
observed in the 1d DT model is very long. It is interesting to note in this
context that recent simulations of the DT model in two dimensions \cite{patcha}
show that this system also exhibits transient multiscaling, but over a much
shorter interval of time. This may be related to the fact that terms involving
higher powers of the height gradient are marginally 
irrelevant in the 2d LD equation. 
Further
investigation of the role of these higher-order terms in the behavior of 
growth equations would be interesting and important for a complete 
understanding of the problem.

This work is supported by the US-ONR and the NSF-DMR-MRG. 
CD and JMK would like to thank the Condensed Matter
Physics Group of the University of Maryland for hospitality. One of the 
authors (JMK) also wishes to thank Hallym Academy of Science, Hallym University
for support.

\begin{figure}
\caption{The rms interface width $W$ and the moments $\sigma_q$, $q$ = 1,4,
of the nearest-neighbor height difference (see text) as functions of time
$\tau$ for the 1d discretized LD equation with $\lambda$ = 1.0. The data shown
were obtained for a system with $L$ = 10$^4$, using an integration time step 
$\Delta \tau$ = 0.01. Inset: The height-difference
correlation functions $G_q(l)$, $q$ = 1,4, (see text)
as functions of the separation
$l$ for this system at time $\tau$ = 10$^4$.}
\label{fig1}
\end{figure}

\begin{figure}
\caption{The rms interface width $W$ and the moments $\sigma_q$, $q$ = 1,4,
of the nearest-neighbor height difference as functions of time
$\tau$ for the 1d KD model with $\lambda$ = 0.5. The data shown represent an
average over 10 runs on systems with $L$ = 10$^4$. 
Inset: The height-difference
correlation functions $G_q(l)$, $q$ = 1,4, as functions of the separation
$l$ for this system at time $\tau$ = 10$^4$.}
\label{fig2}
\end{figure}

\begin{figure}
\caption{The rms interface width $W$ and the moments $\sigma_q$, $q$ = 1,4,
of the nearest-neighbor height difference as functions of time
$\tau$ for the 1d KD model with $\lambda$ = 1.0. The data shown represent an
average over 10 runs on systems with $L$ = 10$^4$. }
\label{fig3}
\end{figure}

\begin{figure}
\caption{ Growth probability (see text) of a pillar 
in the 1d discretized LD equation ($\lambda$ = 1.0, $\Delta \tau$ = 0.01, 
2000 runs on $L$ = 100 samples)
as a function of time $\tau$. Results for three different values,
14, 17 and 20, of the initial height $h_0$ are shown. }
\label{fig4}
\end{figure}

\begin{figure}
\caption{ Growth probability of a pillar of initial height $h_0$ in the 1d
discretized LD equation ($\lambda$ = 1.0, 2000 runs on $L$ = 100 samples) at
time $\tau$ = 1 as a function of $h_0$,
calculated with three different values, 0.01, 0.001 and
0.0001, of the integration time step $\Delta \tau$.}
\label{fig5}
\end{figure}

\begin{figure}
\caption{The dependence of $h_c$, the critical height of a pillar (or 
the depth of a groove, see text), on
the coupling constant $\lambda$ in the discretized LD equation in one and two
dimensions and the discretized KPZ equation in one dimension. The results
shown were obtained from numerical integrations with $\Delta \tau$ = 0.01.
The solid lines are the best fits to the form $h_c(\lambda) = A/\lambda$. }
\label{fig6}
\end{figure}

\begin{figure}
\caption{Development of the instability induced 
in the 1d discretized LD equation ($\lambda$ = 1.0)
by the presence of a pillar of initial height 
$h_0$ = 25. Interface profiles at
times $\tau$ = 0.05, 0.1, 0.15 and 0.17, obtained for a $L$ = 100 system using
$\Delta \tau$ = 10$^{-4}$, are shown. In the initial state, the height is 
zero everywhere except at the 50th site where the height is $h_0$.}
\label{fig7}
\end{figure}

\begin{figure}
\caption{Development of the instability induced 
in the 1d discretized KPZ equation ($\lambda$ = 1.0)
by the presence of a groove of initial depth 
$h_0$ = 30. Interface profiles at
times $\tau$ = 0.1, 0.3 and 0.5, obtained for a $L$ = 100 system using
$\Delta \tau$ = 10$^{-4}$, are shown. In the initial state, the height is 
zero everywhere except at the 50th site where the height is $ - h_0$.}
\label{fig8}
\end{figure}

\begin{figure}
\caption{ Probability of instability (see text)
induced by a groove of initial depth 
$h_0$ in the 1d
discretized KPZ equation ($\lambda$ = 1.0, 2000 runs with $\Delta \tau$ = 0.01
on $L$ = 100 samples) as
a function of time $\tau$.
Results for three different values, 24, 25 and 26, of the initial height
$h_0$ are shown.}
\label{fig9}
\end{figure}

\begin{figure}
\caption{Time
dependence of the maximum nearest-neighbor height difference
$s_{max}$ in the 1d discretized LD equation ($\lambda$ = 4, $\Delta \tau$ = 
0.01, $L$ = 1000, 200 runs),
the 2d discretized LD equation ($\lambda$ = 5, $\Delta \tau$ = 0.01, system
size = 200$\times$200, 30 runs) and the 1d
discretized  KPZ equation
($\lambda$ = 5, $\Delta \tau$ = 0.01, $L$ = 10$^4$, 200 runs). 
The solid line is a fit of the data for the
1d LD equation to the form $s_{max}(\tau) = a + b \ln
\tau$.} 
\label{fig10}
\end{figure}

\begin{figure}
\caption{The average instability time $\tau_{ins}$ (see text) as a function
of $\lambda$ for the 1d discretized LD equation ($L$ = 1000, $\Delta \tau$ =
0.01, number of runs = 150, 200, 500, 1000, 1000 for $\lambda$ = 3, 4, 5,
6, 7, respectively) and the 1d discretized KPZ
equation ($L$ = 10$^4$, $\Delta \tau$ = 0.01,  200 runs for each value of
$\lambda$). The arrow indicates a lower bound for
$\tau_{ins}$ for the KPZ equation with $\lambda$ = 4. The solid line
is a fit of the LD equation data for $\lambda$ = 4, 5 and 6 
to the form $\tau_{ins} = A e^{B/\lambda^2}$. }
\label{fig11}
\end{figure}

\begin{figure}
\caption{The ratios $\sigma_q(\tau)/\sigma_1(\tau)$, $q$ = 2, 3 ,4 and
5, as functions of time $\tau$ for the 1d discretized LD equation with 
$\lambda$ = 4.0, $L$ = 1000, $\Delta \tau$ = 0.01, averaged over 100 runs.
The observed growth of the values of these ratios with time indicates 
deviations from single-exponent scaling.}
\label{fig12}
\end{figure}

\begin{figure}
\caption{The dependence of the quantity $\Delta$, which measures the 
initial rate of
growth of the height of an isolated pillar in the 1d CLD model
without noise, on $h_0$, the initial height of the pillar. The results
shown are for $\lambda$ = 4.0 and three different values, 0.02, 0.03, and
0.3, of the control parameter $c$ (see text).}
\label{fig13}
\end{figure}

\begin{figure}
\caption{The probability of growth of an isolated pillar in the 1d CKD model
($\lambda$ = 2.0, $c$ = 0.02) as a function of time. These data were 
obtained from 2000 runs on samples with $L$ = 100. Results for three different
values, 5.0, 25.0 and 45.0, of the initial height $h_0$ are shown.}
\label{fig14}
\end{figure}

\begin{figure}
\caption{The rms interface width $W$ 
as a function of time
$\tau$ for the 1d  CLD model ($\lambda$ = 4.0, $L$ = 1000, $\Delta \tau$ =
0.01) with two values
of $c$ (10 runs for $c$ = 0.02 and 40 runs for $c$ = 0.05), and for the
1d CKD model ($\lambda$ = 2.0, $L$ = 1000) with two values of $c$ (200
runs for $c$ = 0.02 and 10 runs for $c$ = 0.005).}
\label{fig15}
\end{figure}

\begin{figure}
\caption{The rms interface width $W$ and the moments $\sigma_q$,
$q$ = 1 - 4, of the nearest-neighbor height difference 
as functions of time $\tau$ for the 1d CKD
model ($\lambda$ = 2, $c$ =
0.02, $L$ = 1000, averaged over 2000 runs).
Inset: The ratios $\sigma_q(\tau)/\sigma_1(\tau)$, $q$ = 2, 3 and
4, as functions of time $\tau$. The data shown in the inset were averaged over 200 runs.}
\label{fig16}
\end{figure}

\begin{figure}
\caption{The height-difference correlation functions $G_q(l)$,
$q$ = 1 - 4, at time $\tau$ = 1000 
as functions of the separation $l$ for the 1d CKD
model ($\lambda$ = 2, $c$ =
0.02, $L$ = 1000, averaged over 2000 runs). 
The solid lines are power-law fits to the 
data for $l \le$
10. Inset: The ratios $G_q(l)/G_1(l)$, $q$ = 2, 3 and 4 as
functions of the separation $l$.}
\label{fig17}
\end{figure}

\begin{figure}
\caption{ Scaling plot for the dependence of the interface width $W$ in the
1d CKD model ($\lambda$ = 2.0, $c$ = 0.02) on time $\tau$ for systems with
different sizes $L$. The data for $L$ = 20, 40 and 80 were obtained by averaging
over 2000, 1000 and 1000 runs, respectively. The values of the exponents used
in this scaling plot are: $\alpha$ = 1.35, $\beta$ = 0.355.}
\label{fig18}
\end{figure}

\begin{figure}
\caption{Distribution of the nearest-neighbor height difference $s$ in the 1d
CKD model ($\lambda$ = 2.0, $c$ = 0.02). Results are shown for $L$ = 1000,
$\tau$ = 1000 (averaged over 2000 runs) and also for $L$ = 80 in the saturation
regime ($5 \times 10^4 \le \tau \le 5 \times 10^5$), averaged over 1000 runs.
The solid lines represent fits of the data to the power-law form, $P(s) 
\propto s^{-\eta}$, with $\eta$ = 3.2 for the $L$ = 1000 data and $\eta$ = 2.5
for the $L$ = 80 data. The inset shows the best stretched exponential fit to 
the $L$ = 80 data for small values of $s$.}
\label{fig19}
\end{figure}

\begin{figure}
\caption{Typical profile of the interface generated at time $\tau$ = 10$^4$
in the 1d DT model with $L$ = 1000. The quantity plotted along
the vertical axis is the 
deviation of the height from the average height which, by
definition, is equal to $\tau$.} 
\label{fig20}
\end{figure}

\begin{figure}
\caption{Typical profile of the interface generated at time $\tau$ = 10$^4$
in the 1d CKD model with $\lambda$ = 2.0, $c$ = 0.02 and 
$L$ = 1000. 
The average height which, by
definition, is equal to $\tau$, has been subtracted from the values plotted
along the vertical axis.} 
\label{fig21}
\end{figure}

\begin{figure}
\caption{The probability of increase of the depth of 
an isolated groove of width $w$ = 1 in the 1d DT model 
as a function of time $\tau$. The data were obtained by averaging over 10$^5$
runs on samples with $L$ = 32. This probability is independent of the initial
depth $h_0$ of the groove for all $h_0 > 10$. Inset: The probability of 
increase of the depth of an isolated groove of width $w$ = 
3 in the 1d DT model. Results
obtained by averaging over 1000
runs on $L$ = 128 samples are shown
for initial depth $h_0 = 30$ and $h_0 = 90$. }
\label{fig22}
\end{figure}

\begin{figure}
\caption{The probability of growth of an isolated pillar in the 1d 
modified KD model (see text) with
$\lambda$ = 2.0, $u$ = 0.06, as a function of time. These data were 
obtained from 2000 runs on samples with $L$ = 100. Results for four different
values, 10, 30, 50 and 70, of the initial height $h_0$ are shown.}
\label{fig23}
\end{figure}

\begin{figure}
\caption{The rms interface width $W$ and the moments $\sigma_q$,
$q$ = 1 - 4, of the nearest-neighbor height difference 
as functions of time $\tau$ for the 1d  modified KD
model ($\lambda$ = 2, $u$ =
0.06, $L$ = 1000, averaged over 200 runs).
Inset: The ratios $\sigma_q(\tau)/\sigma_1(\tau)$, $q$ = 2, 3 and
4, as functions of time $\tau$.}
\label{fig24}
\end{figure}

\begin{figure}
\caption{The height-difference correlation functions $G_q(l)$,
$q$ = 1 - 4, at time $\tau$ = 10$^4$ 
as functions of the separation $l$ for the 1d modified KD
model ($\lambda$ = 2, $u$ =
0.06, $L$ = 1000, averaged over 200 runs). }
\label{fig25}
\end{figure}

\begin{figure}
\caption{Typical profile of the interface generated at time $\tau$ = 10$^4$
in the 1d  modified KD model with $\lambda$ = 2.0, $u$ = 0.06 and 
$L$ = 1000. The quantity plotted along the $y$-axis is the 
deviation of the height from the average height which, by
definition, is equal to $\tau$.} 
\label{fig26}
\end{figure}
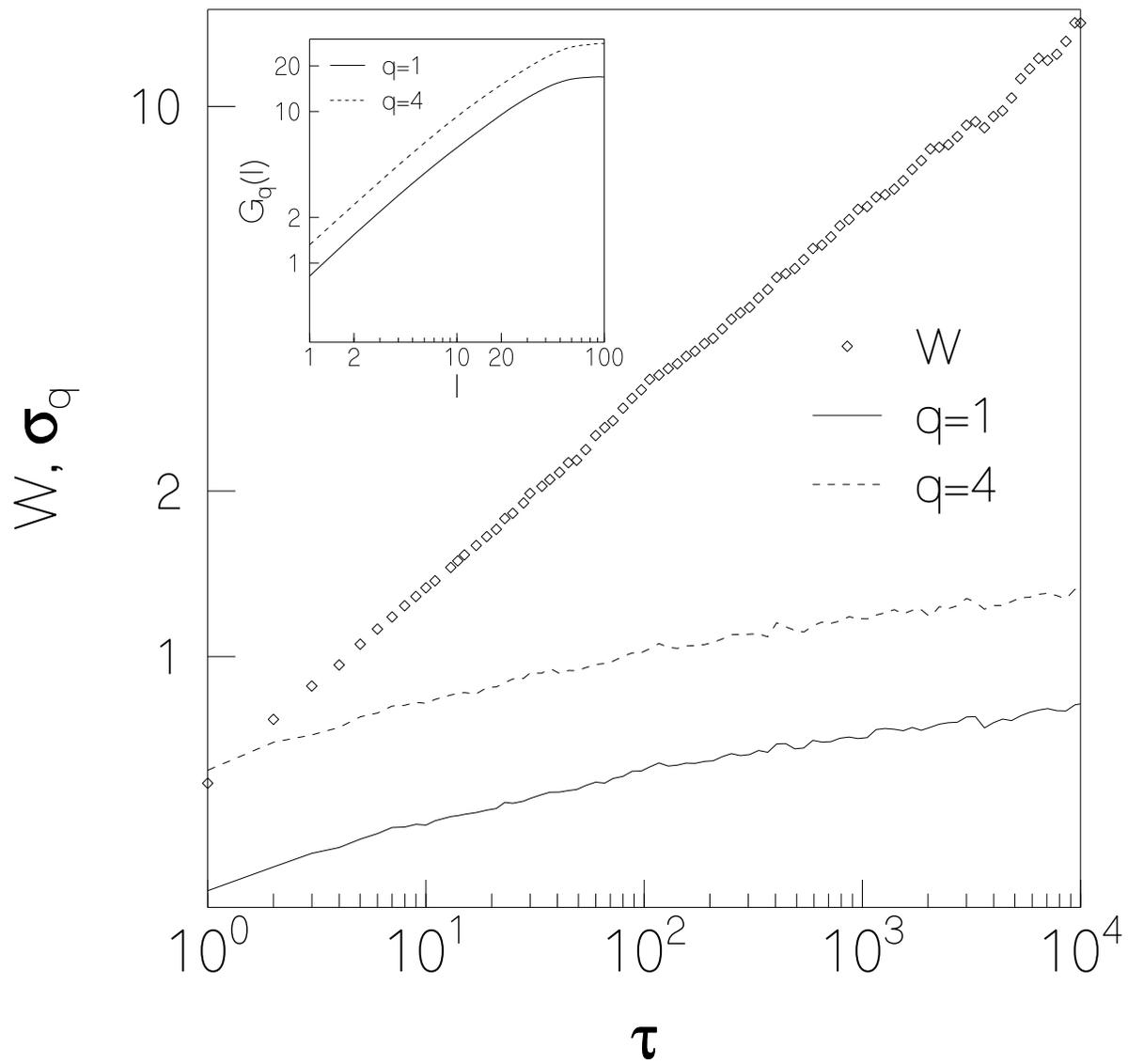
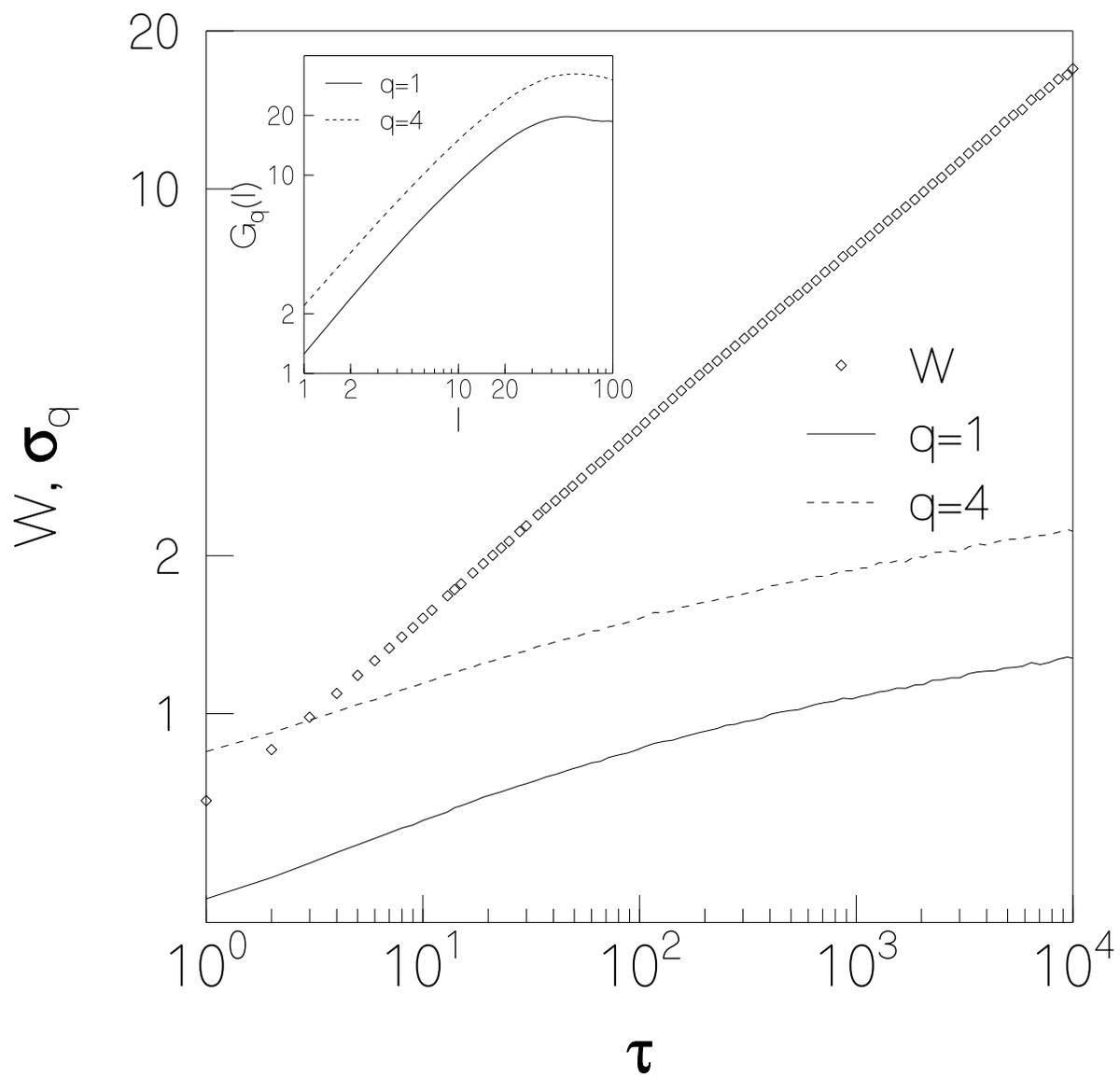
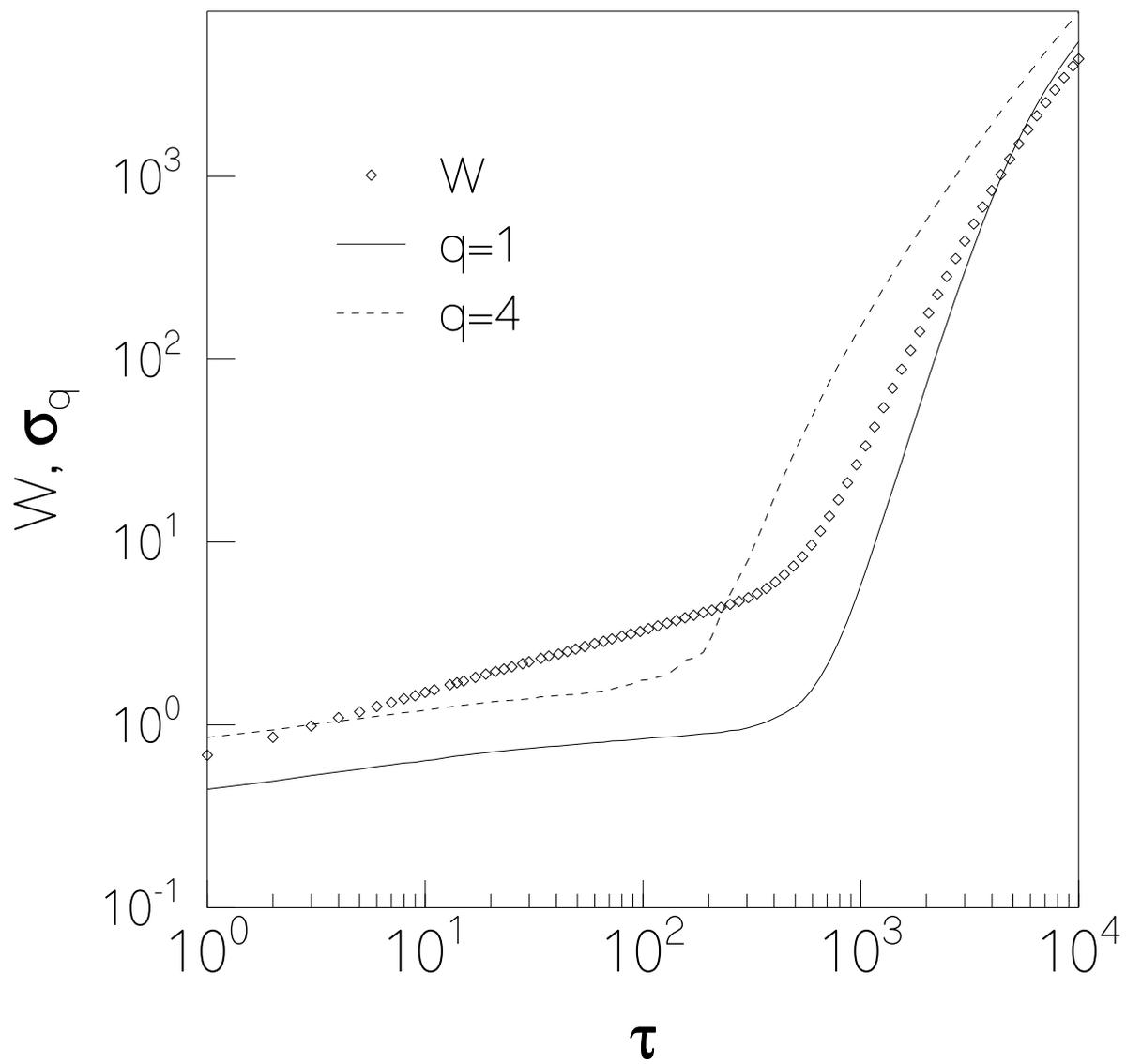
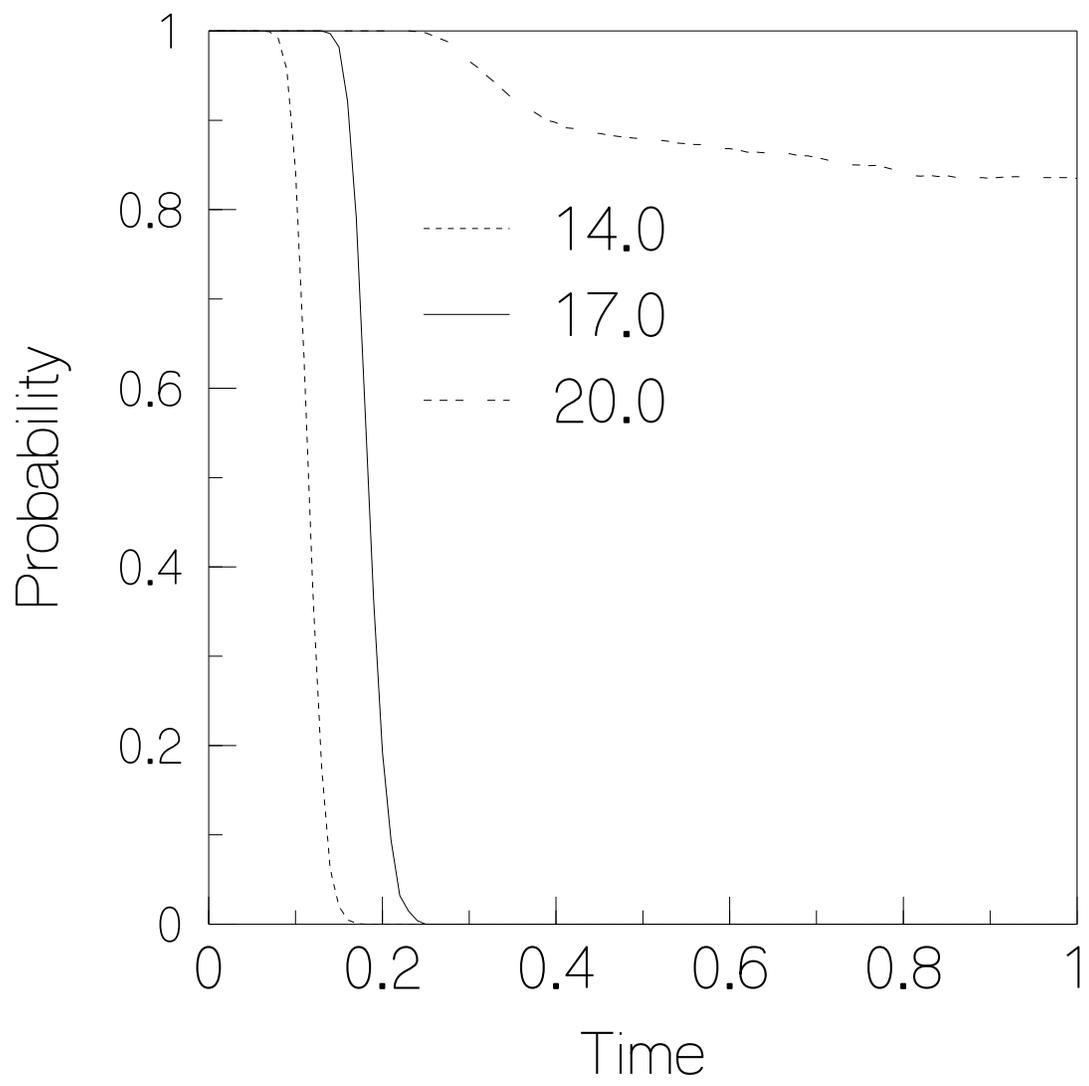
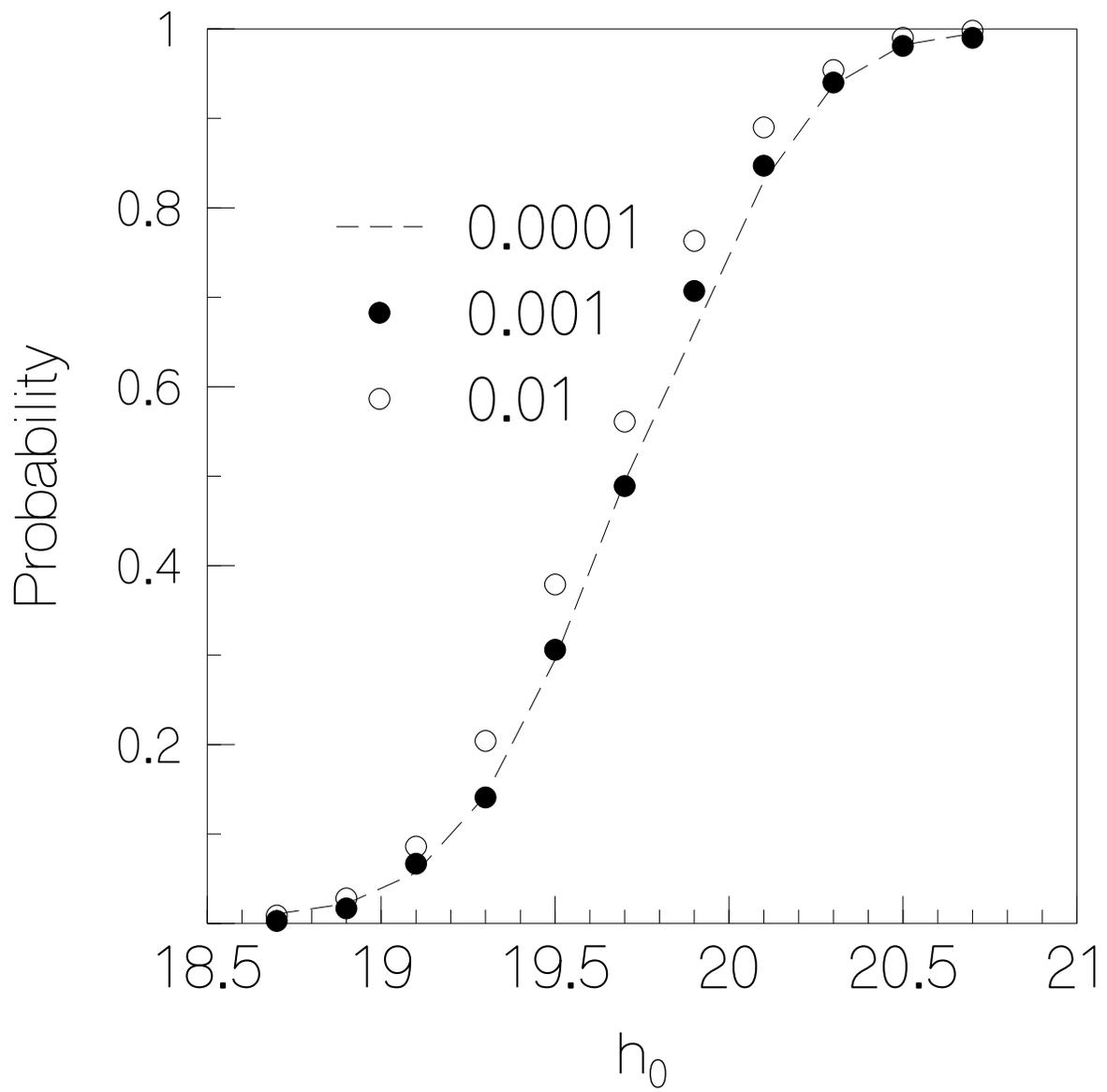
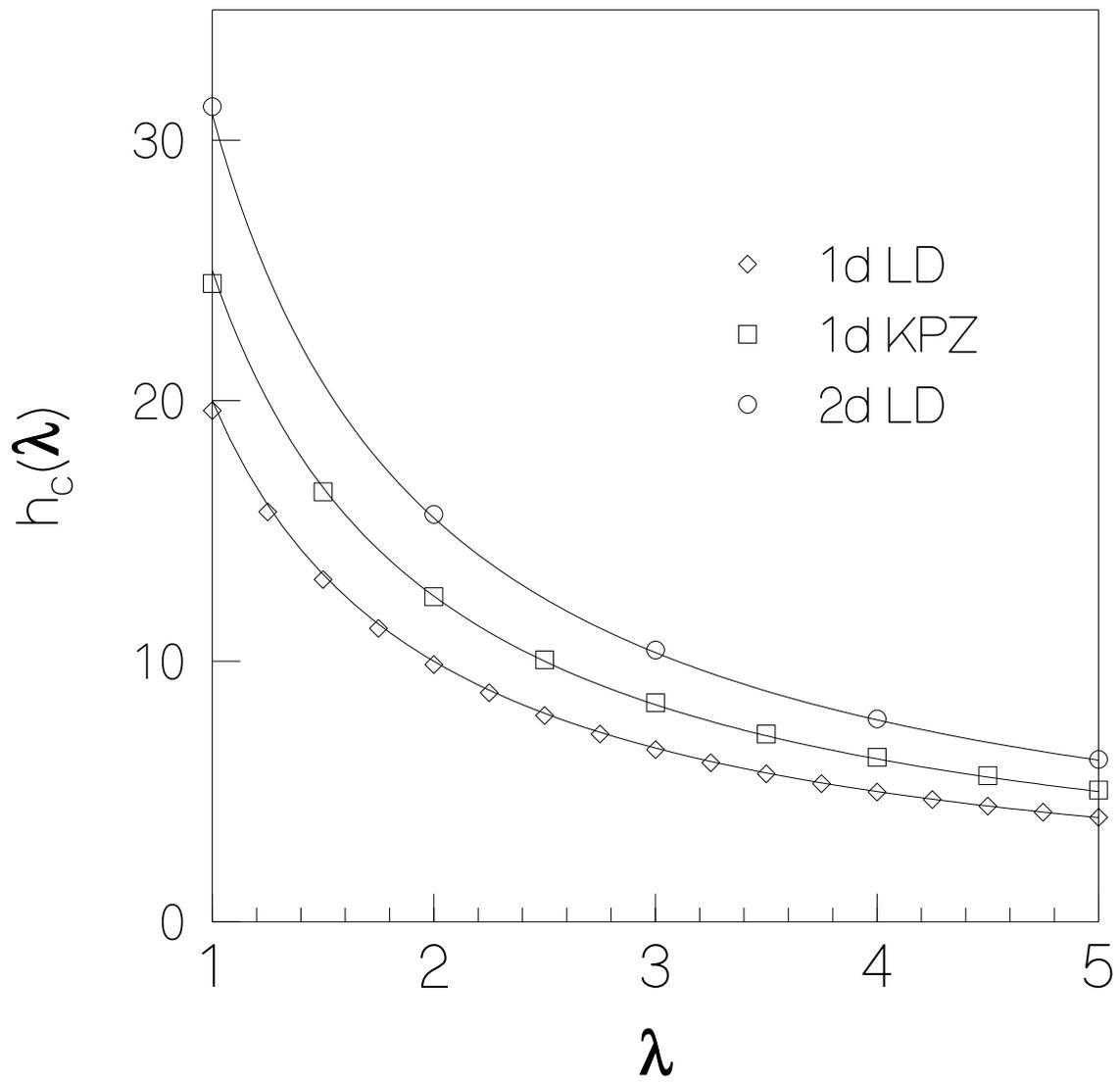
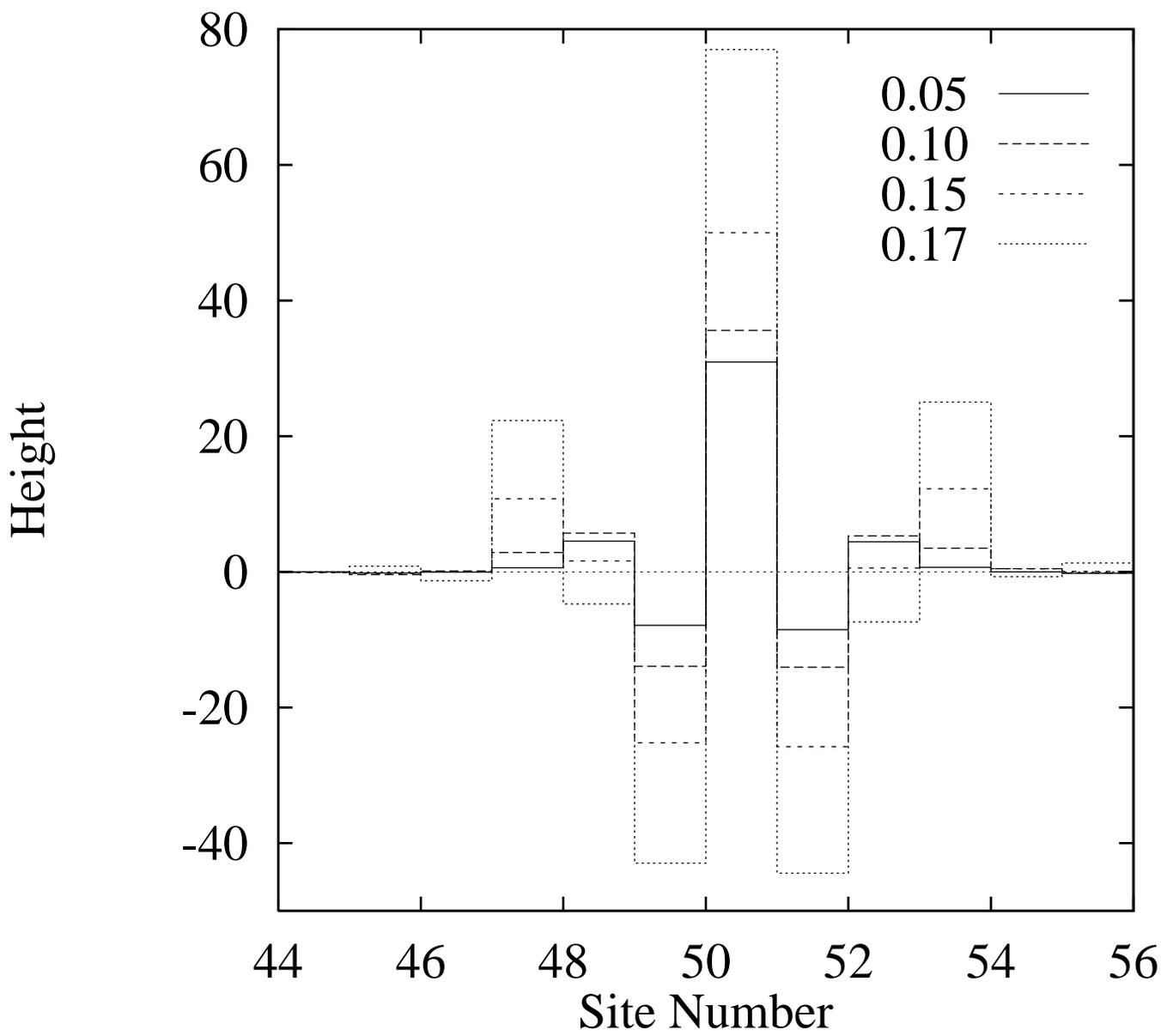
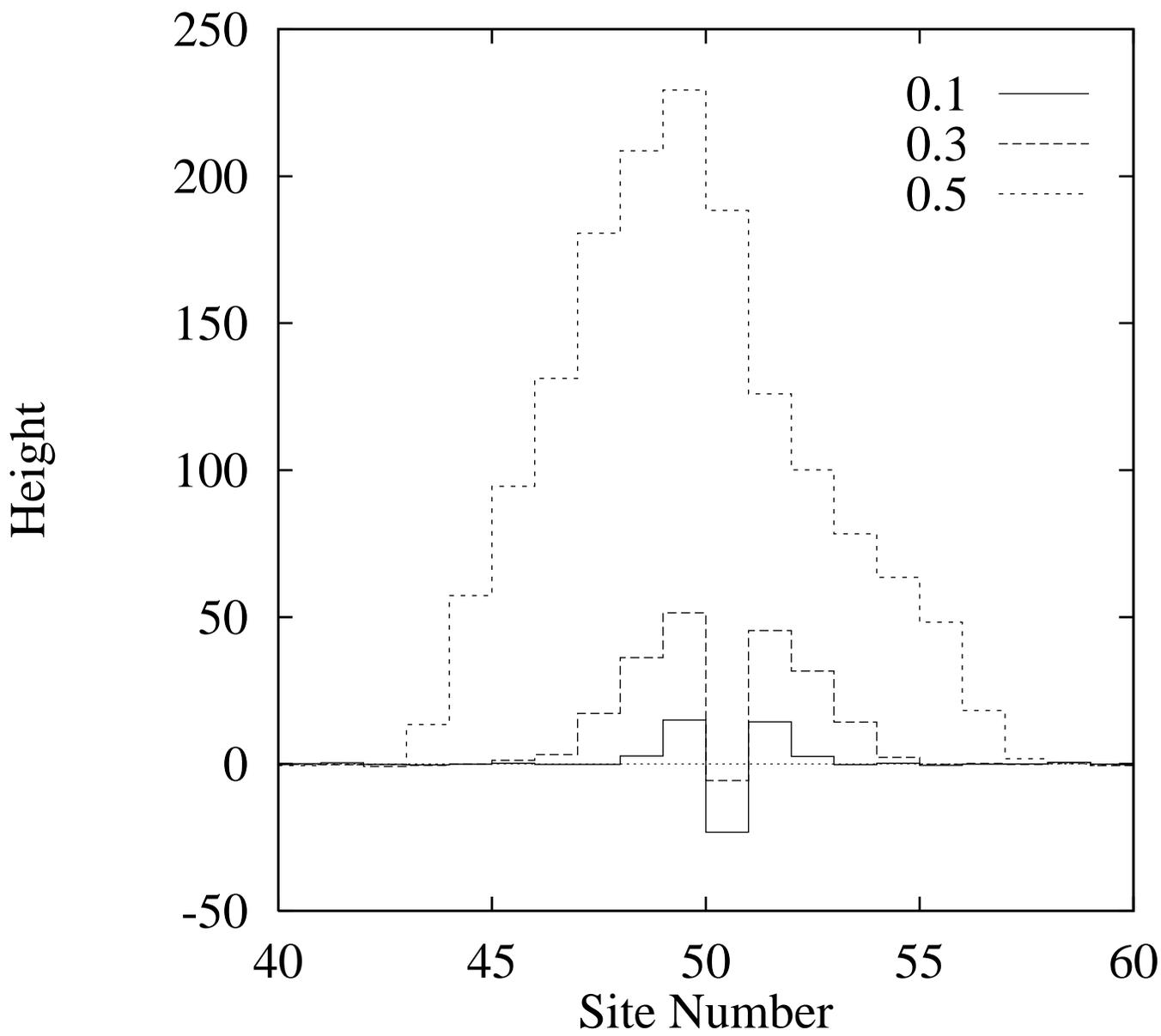
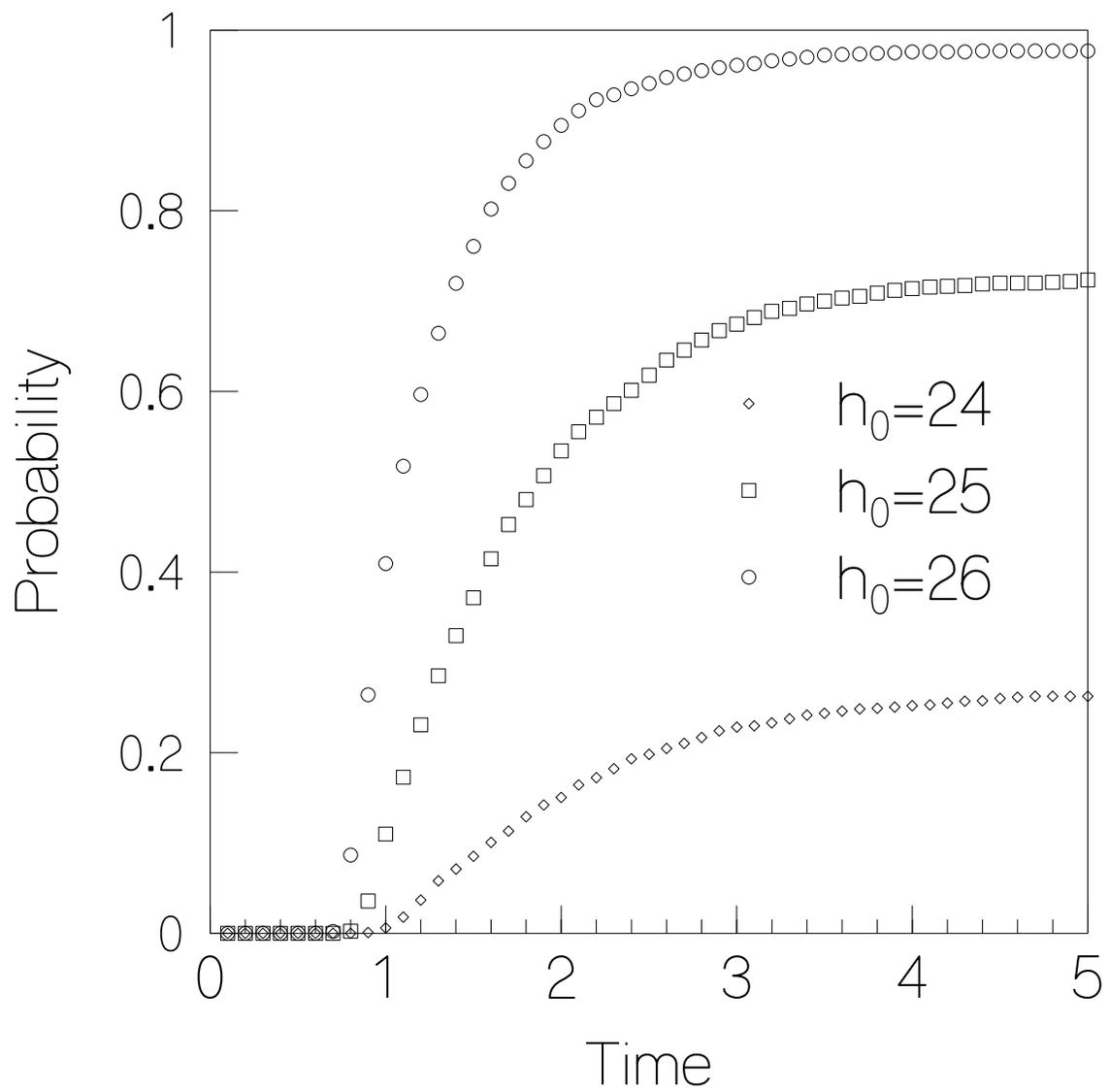
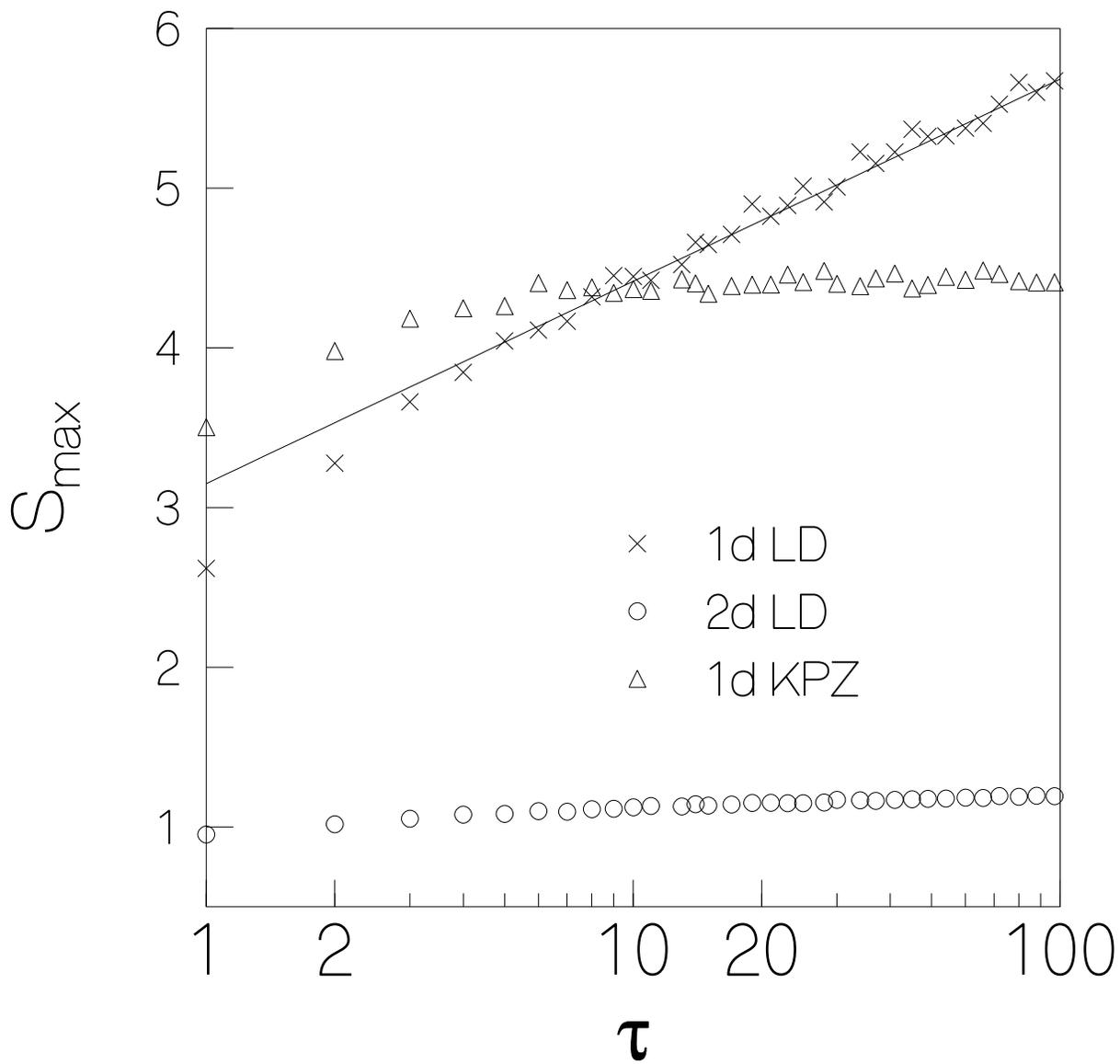
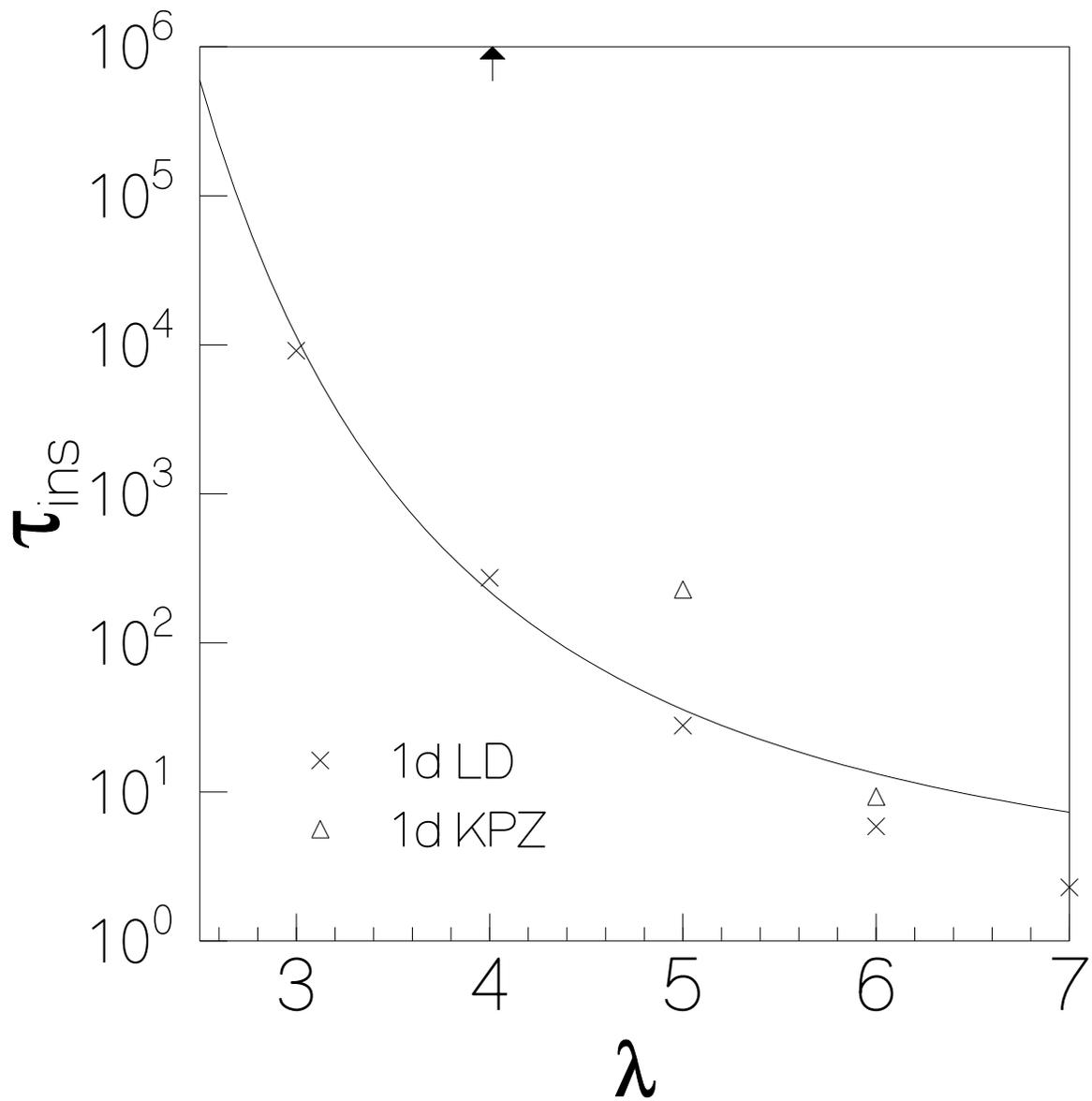
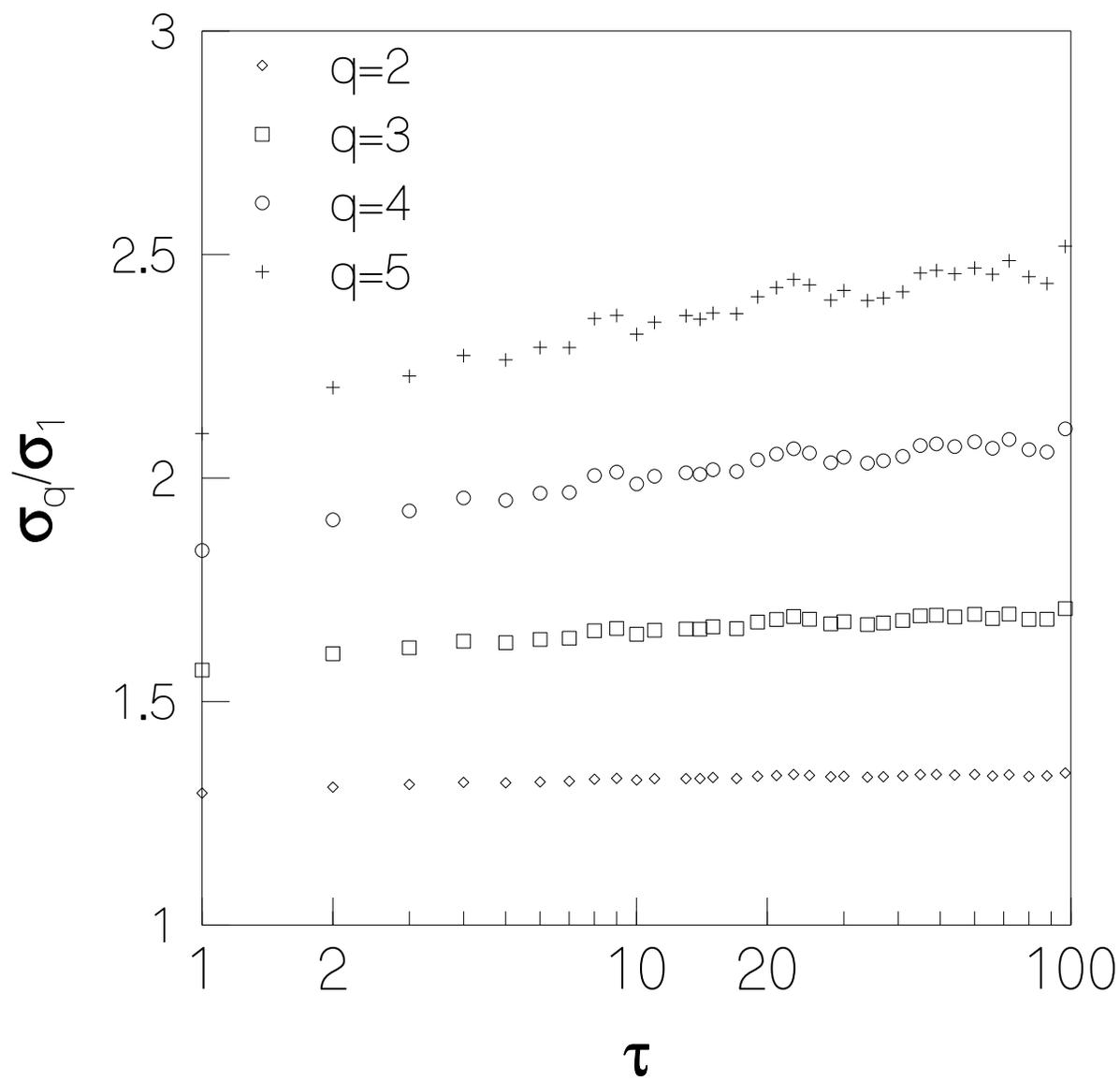
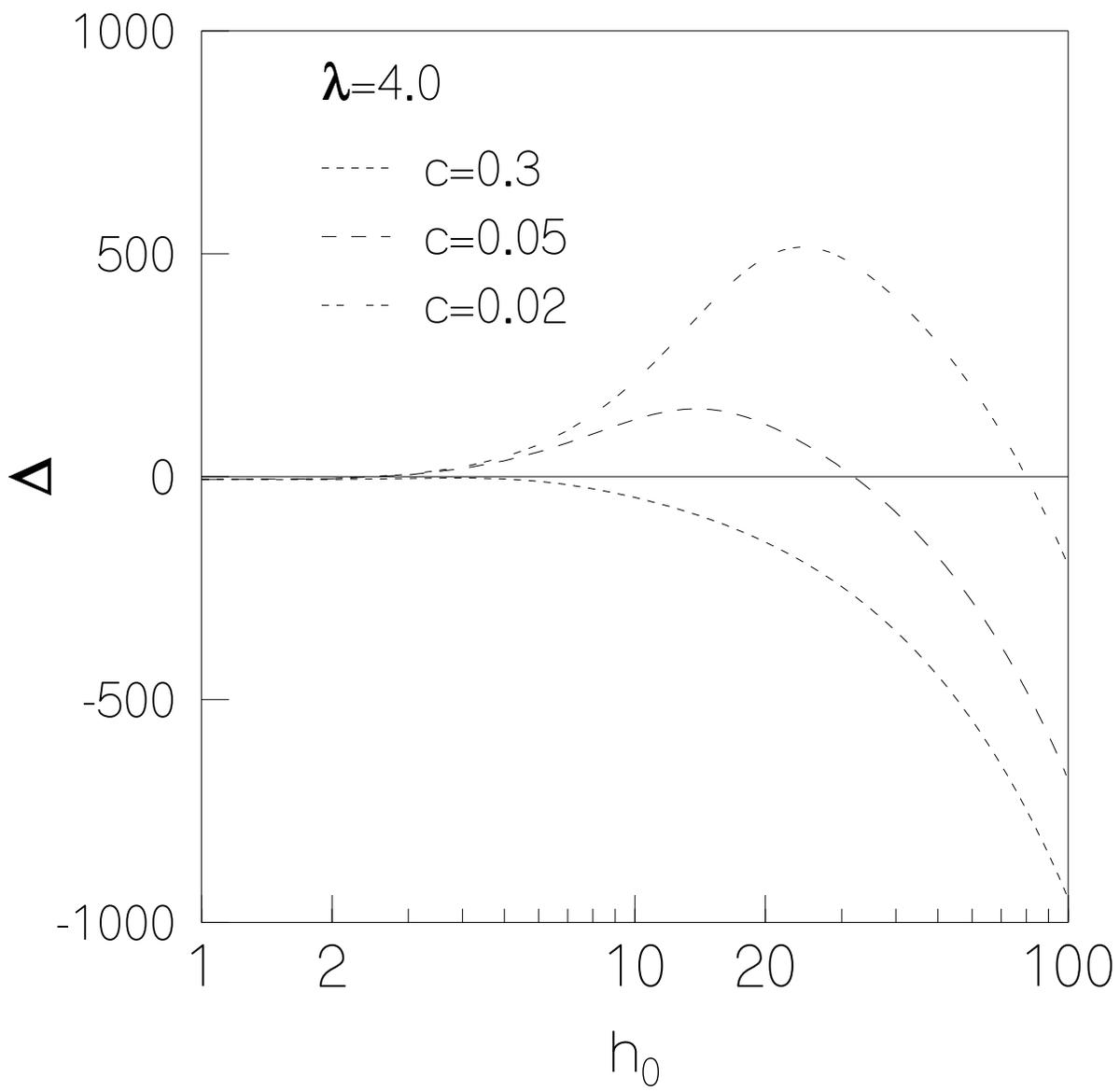
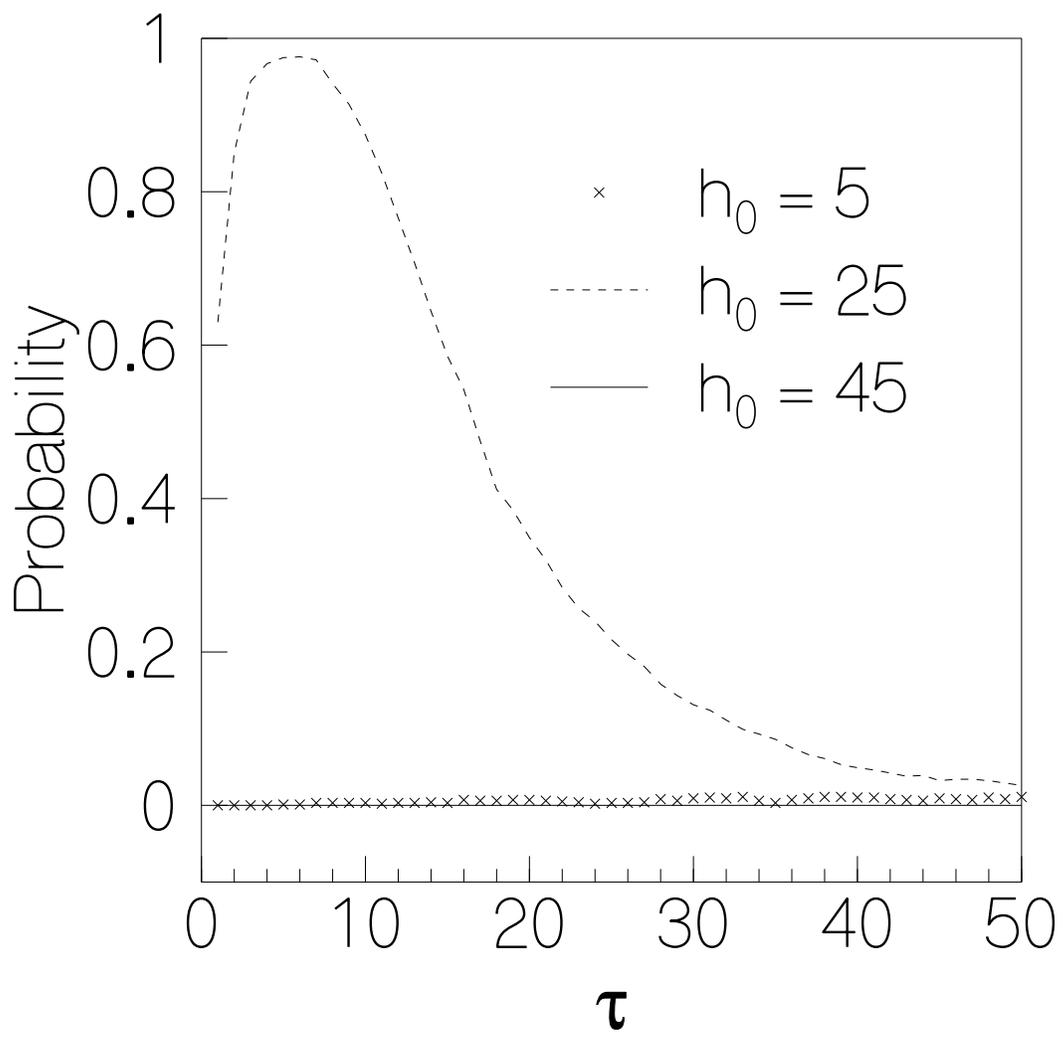
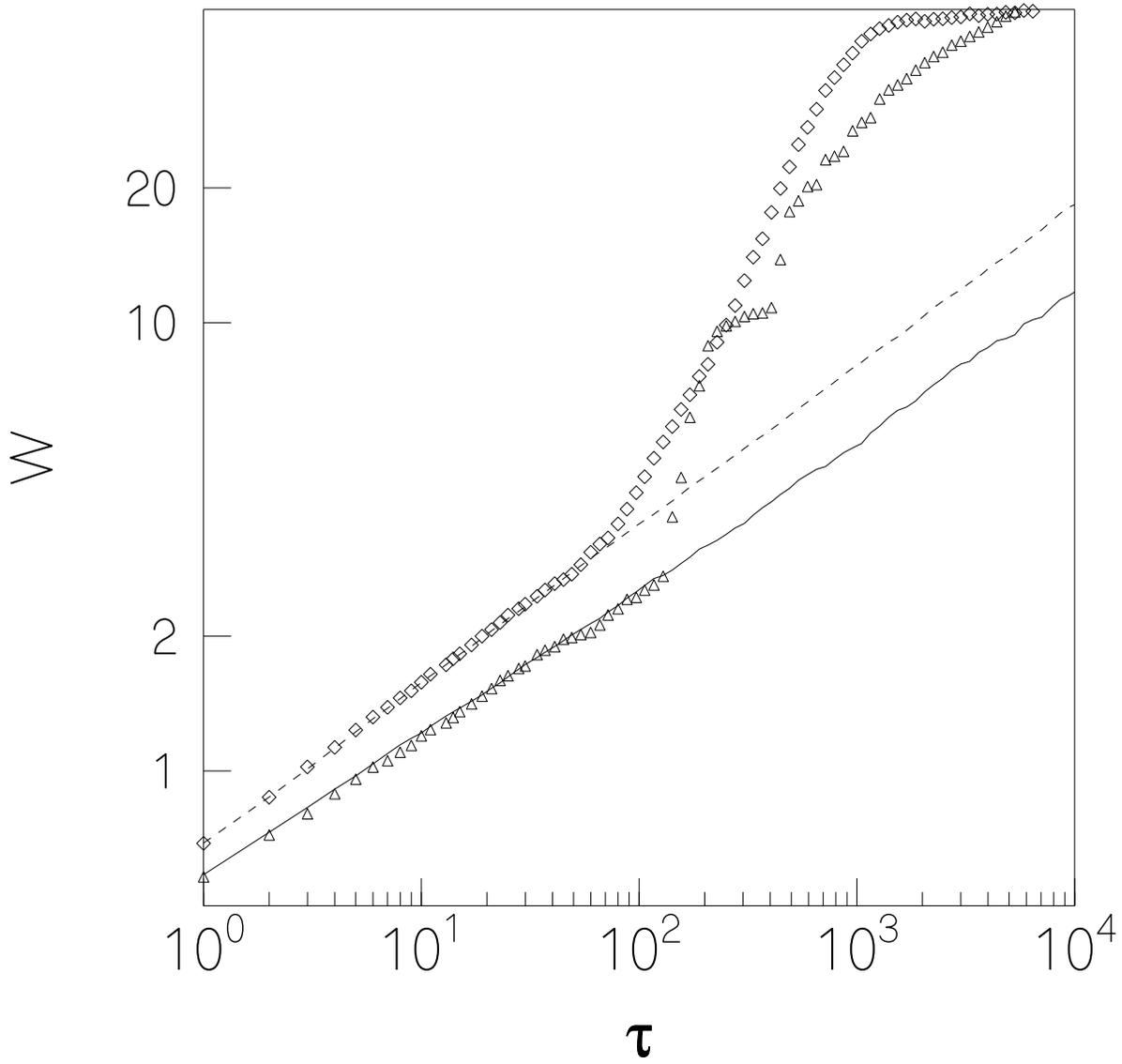
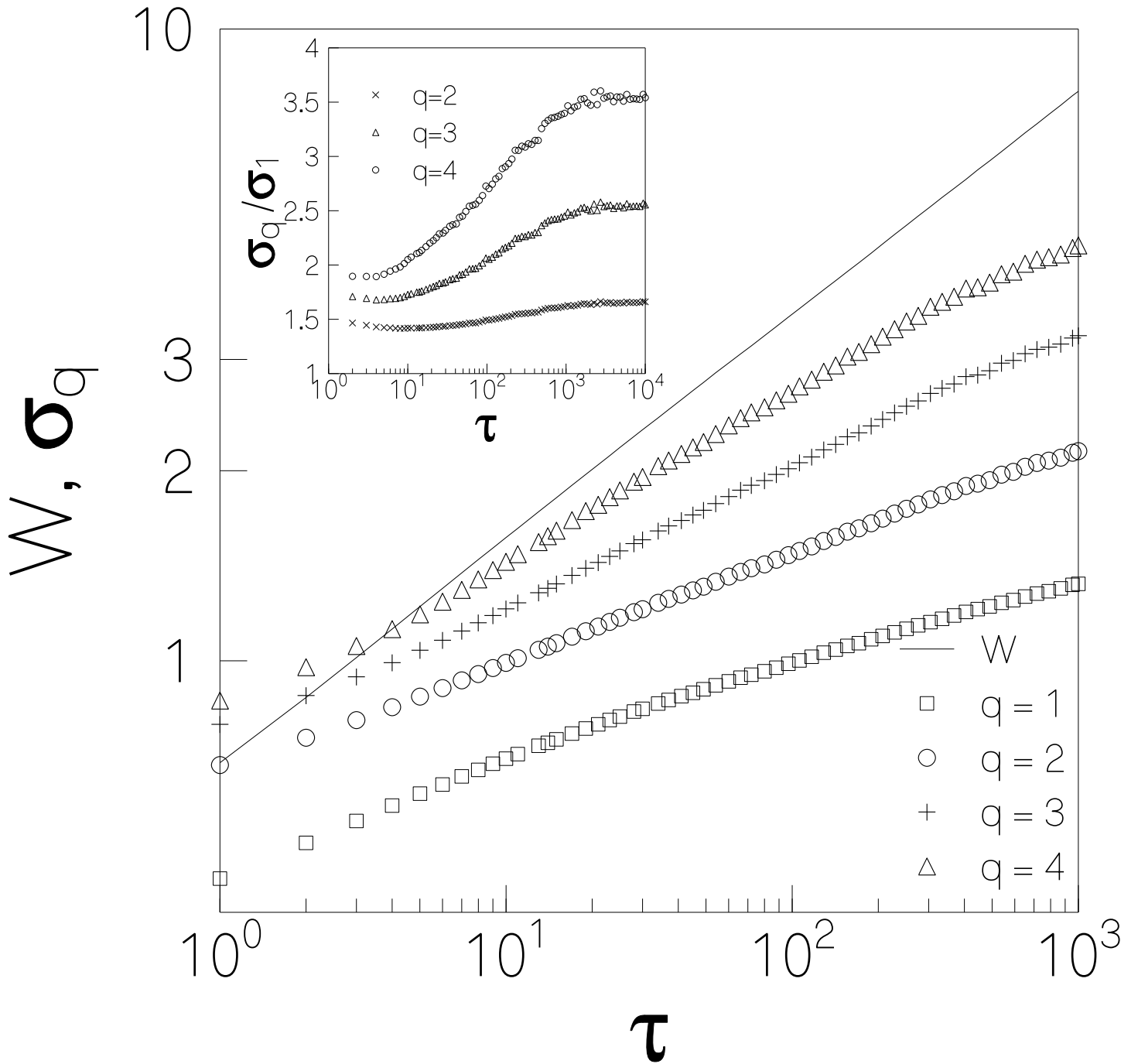
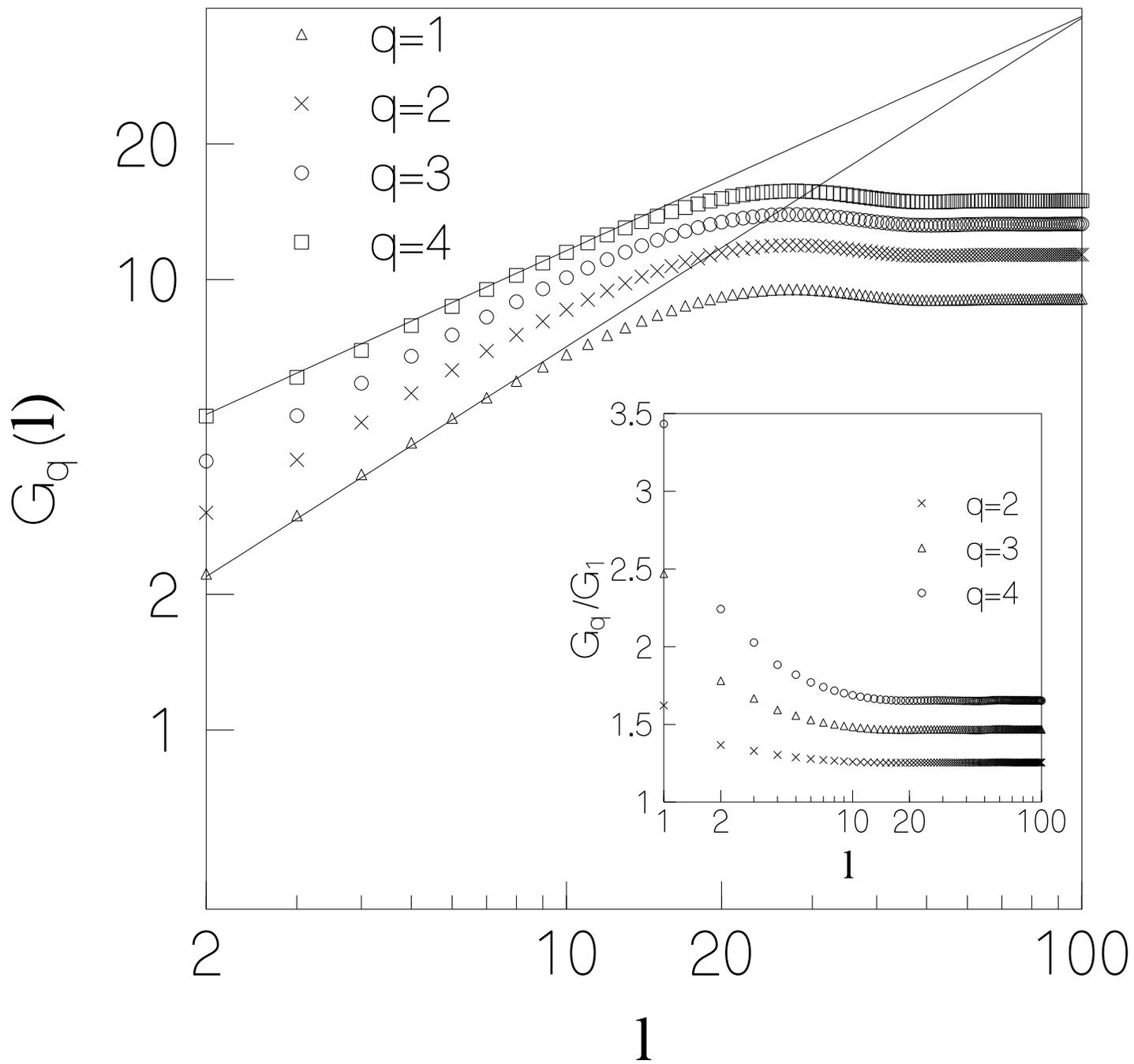
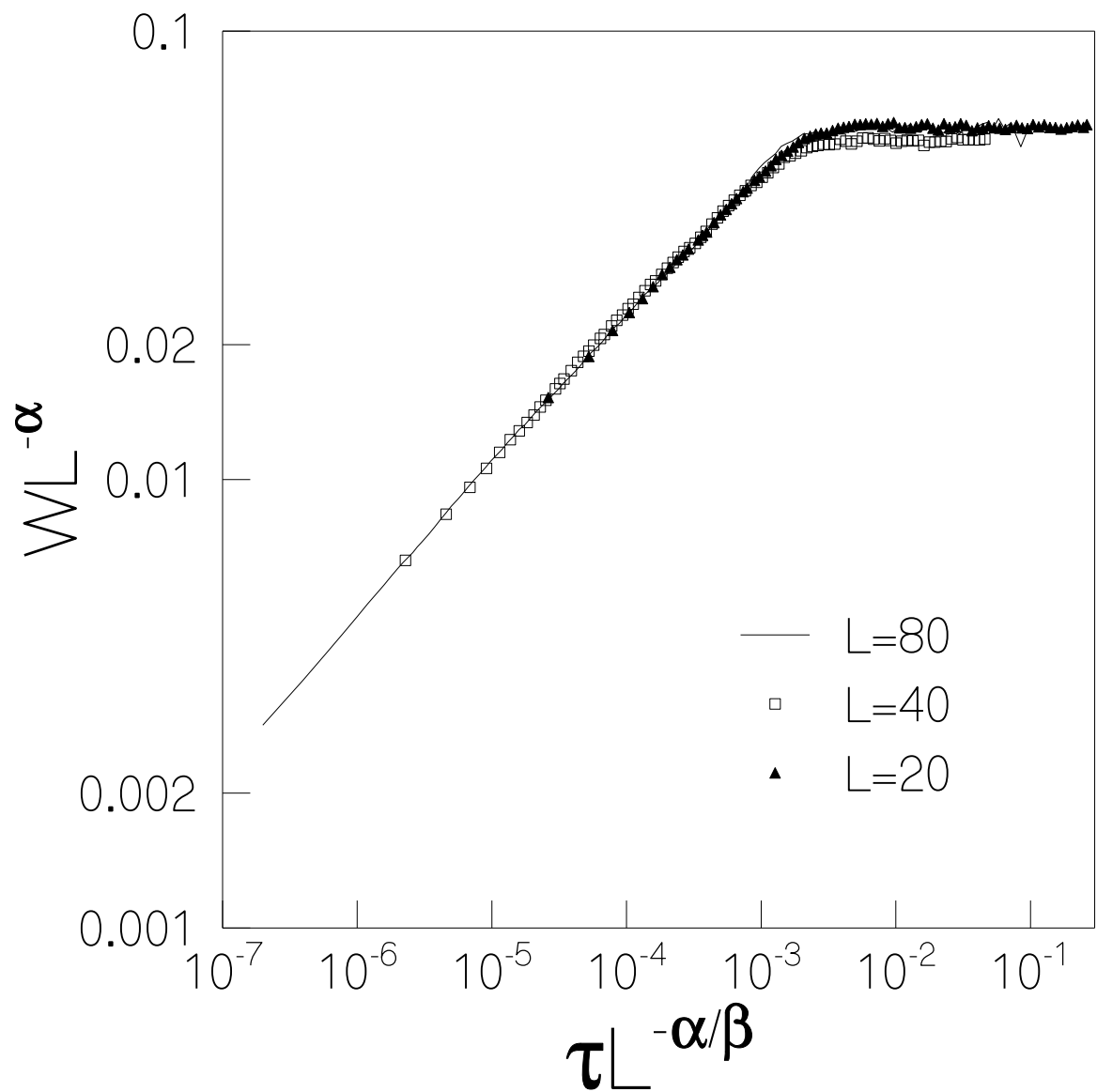
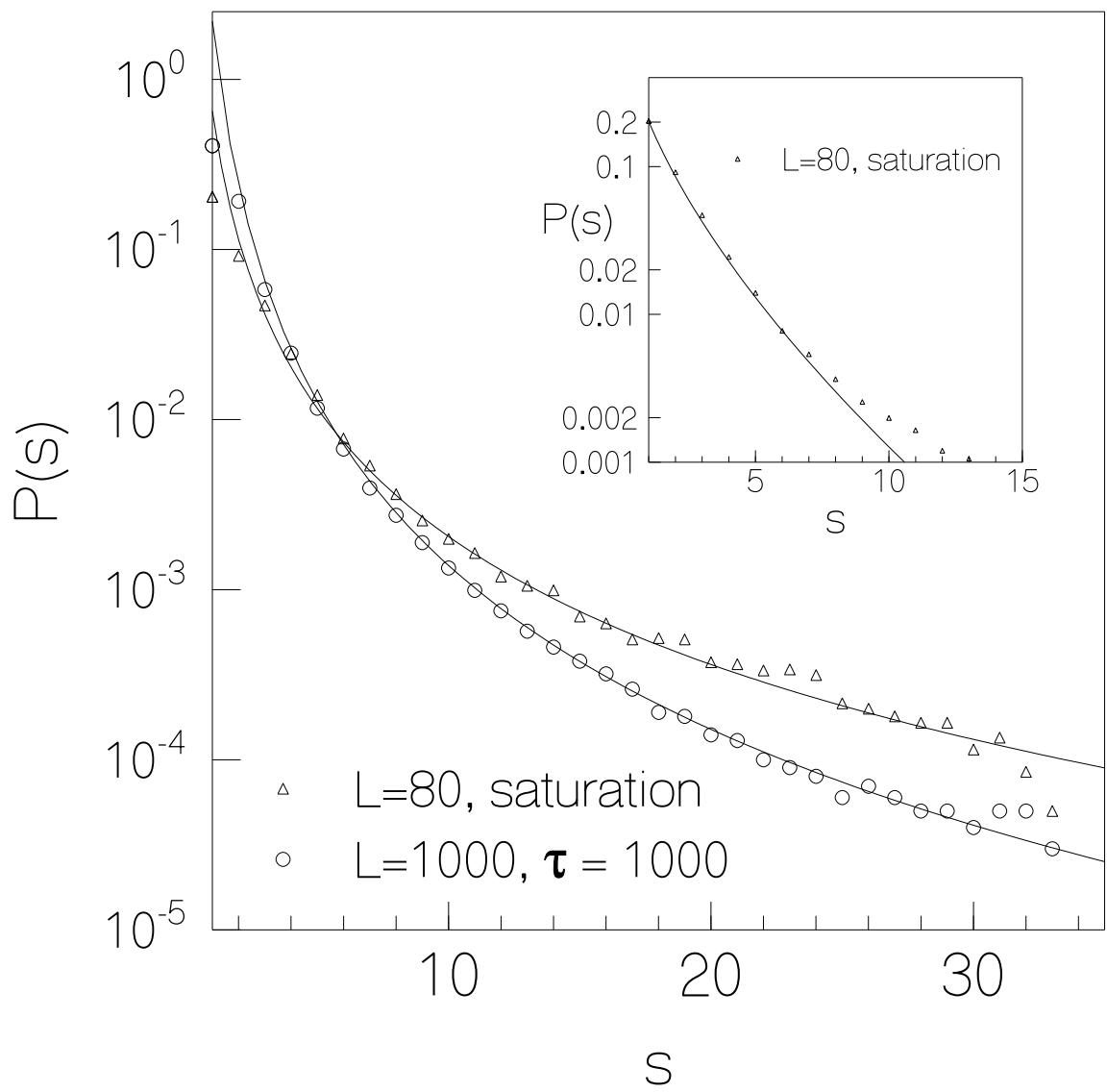
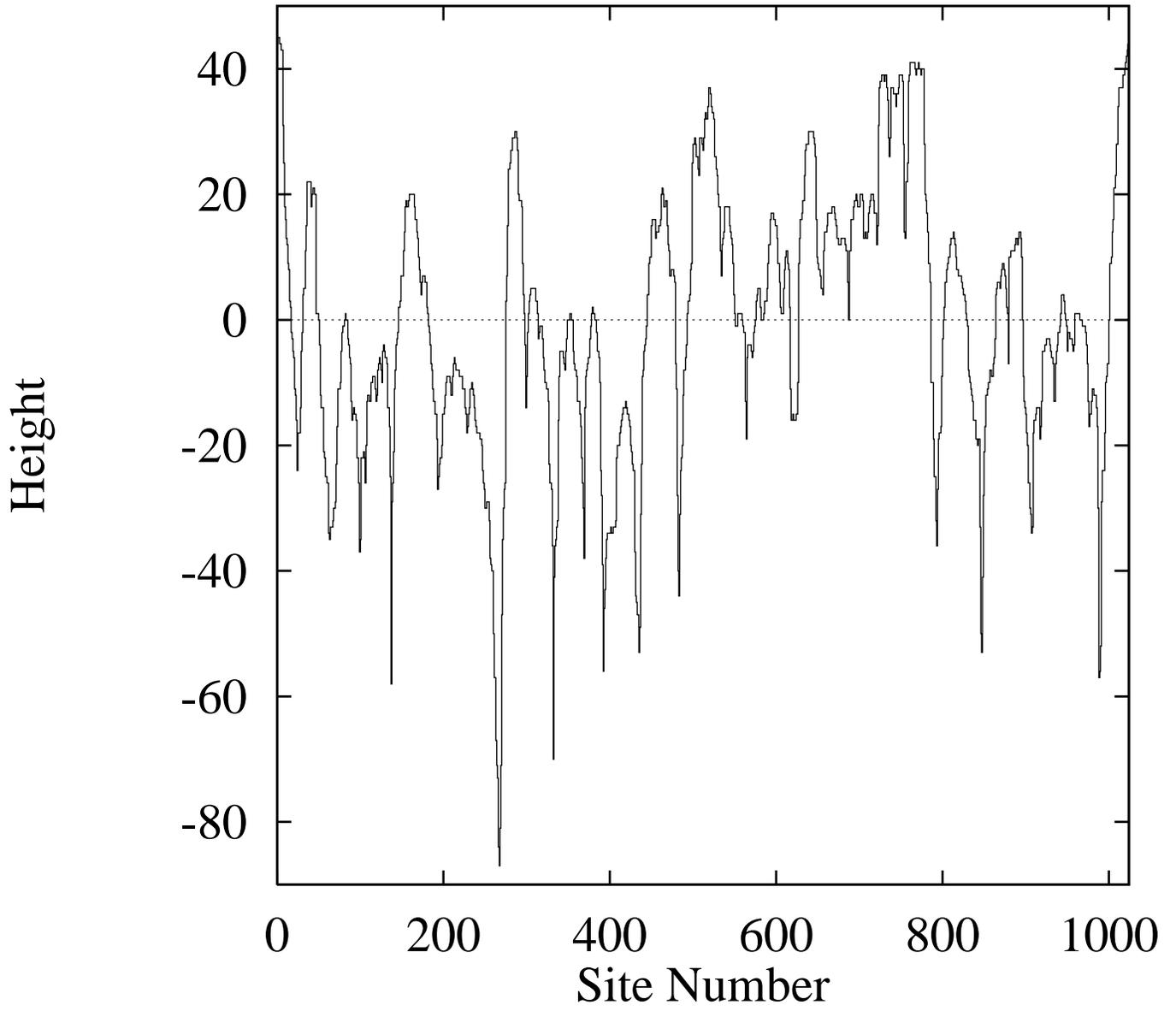
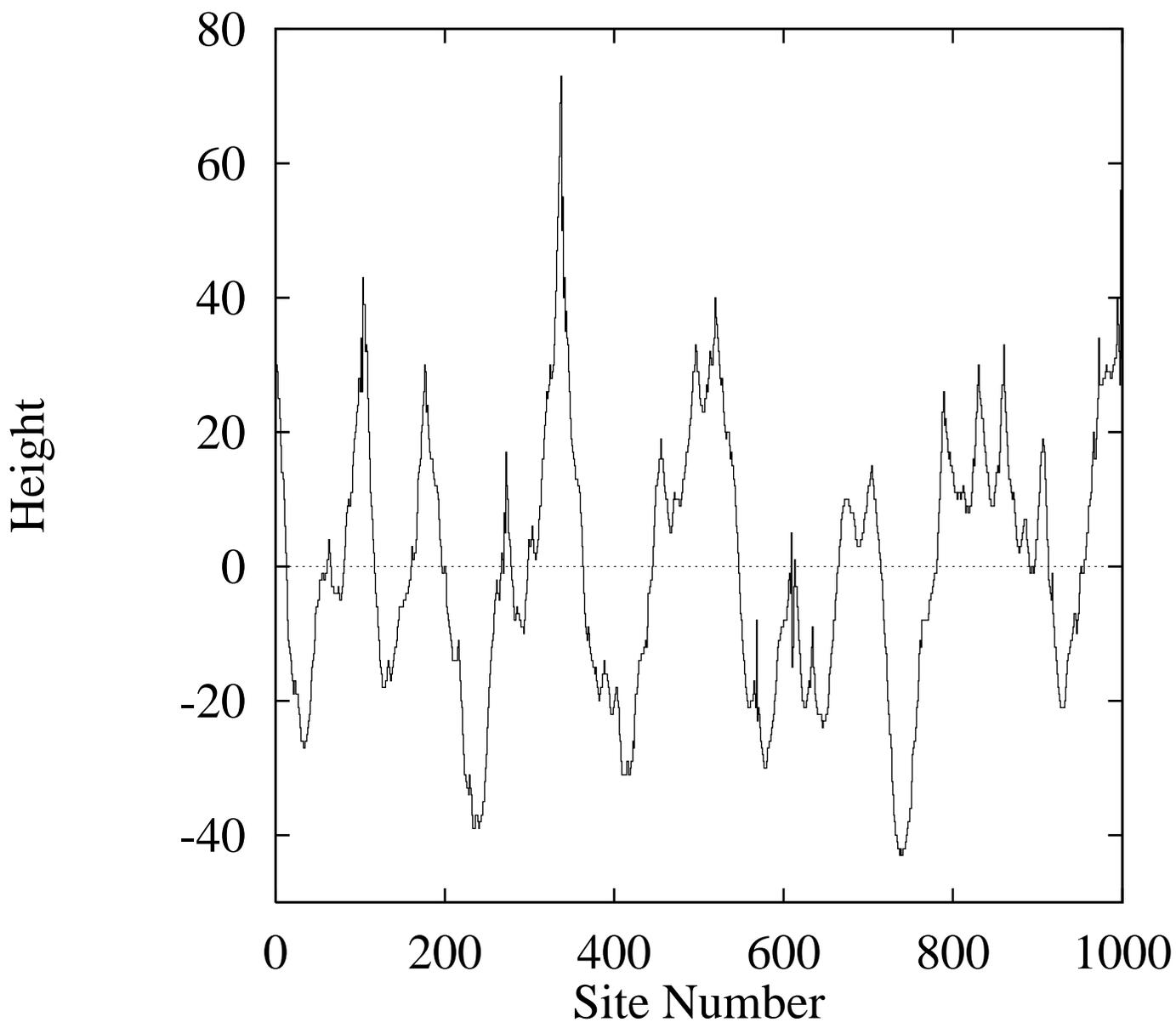
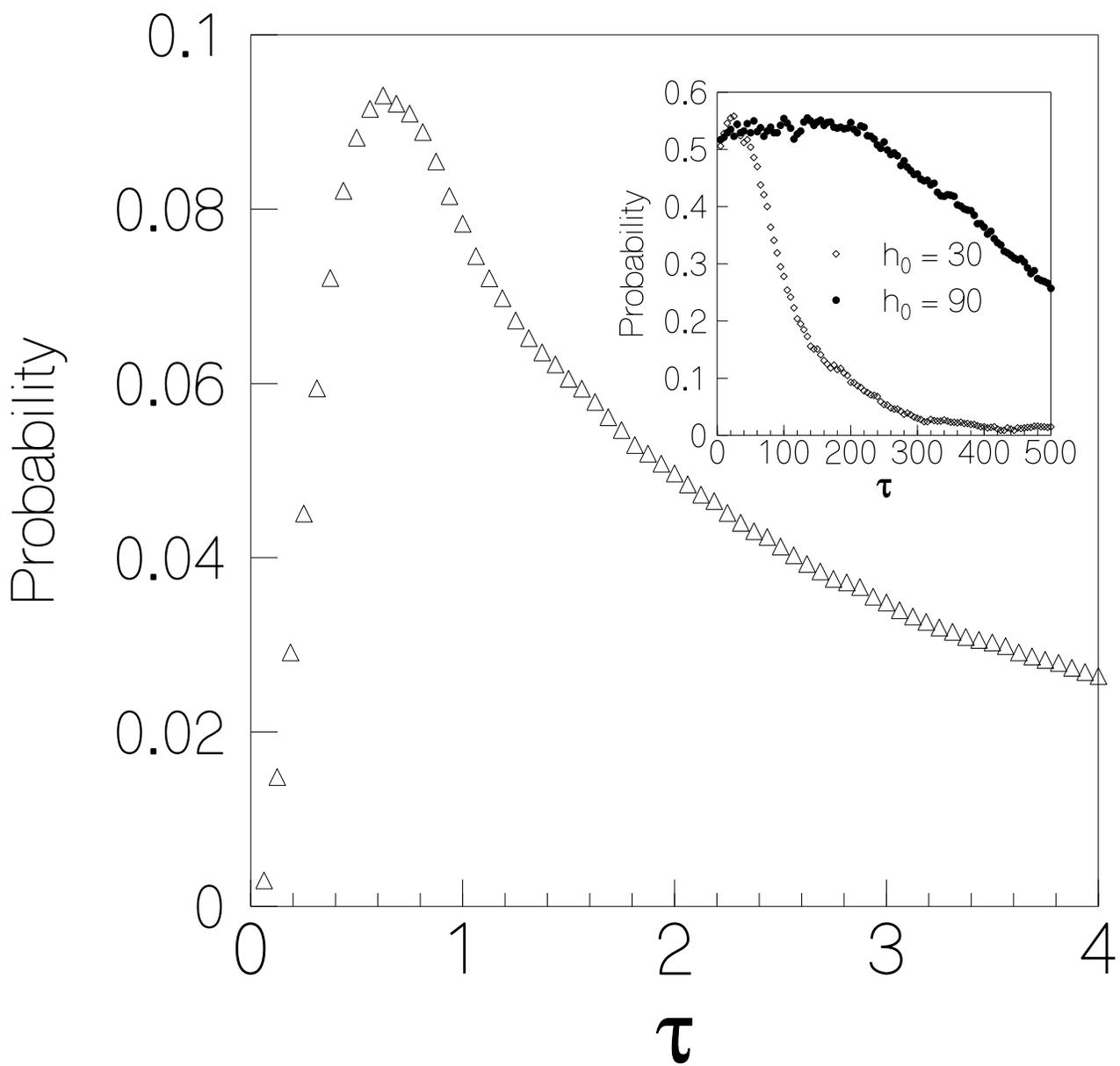
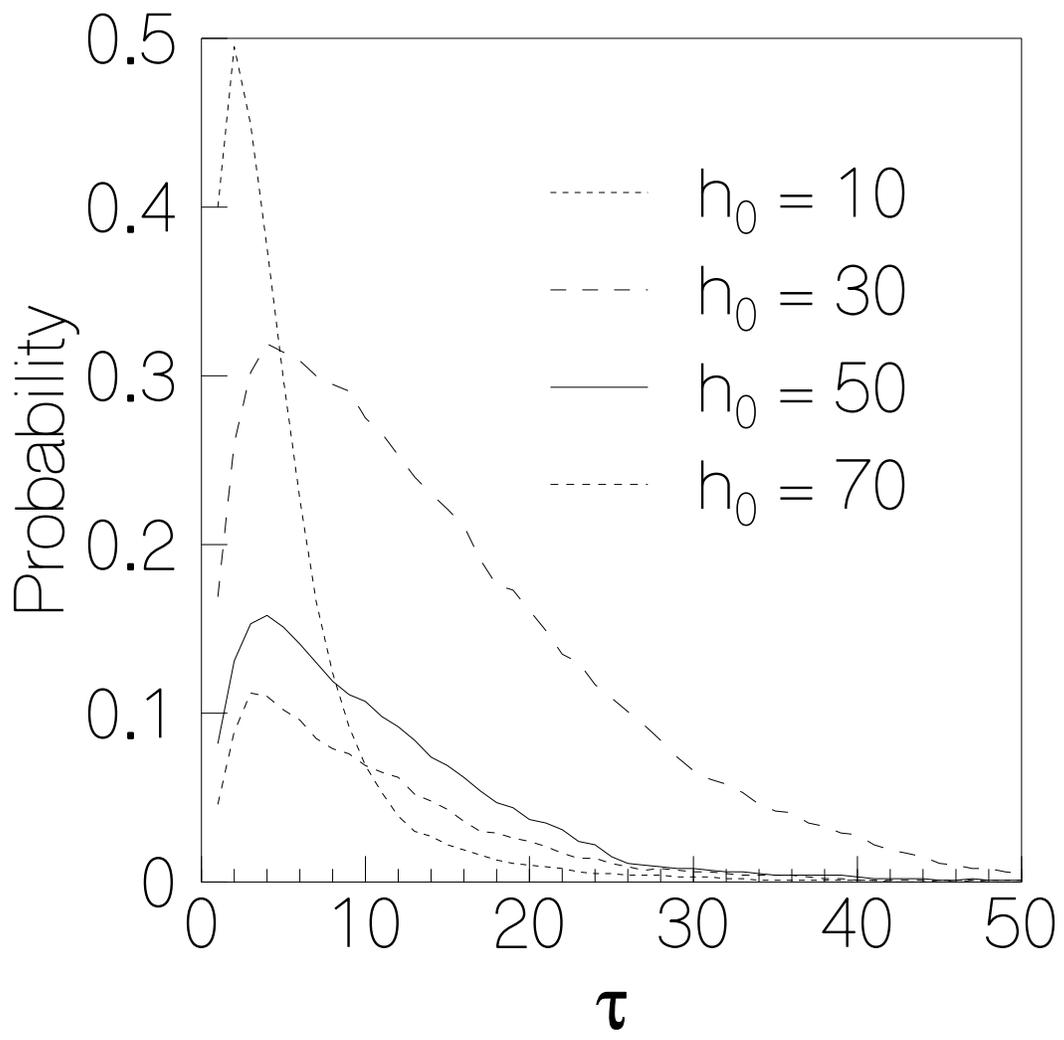
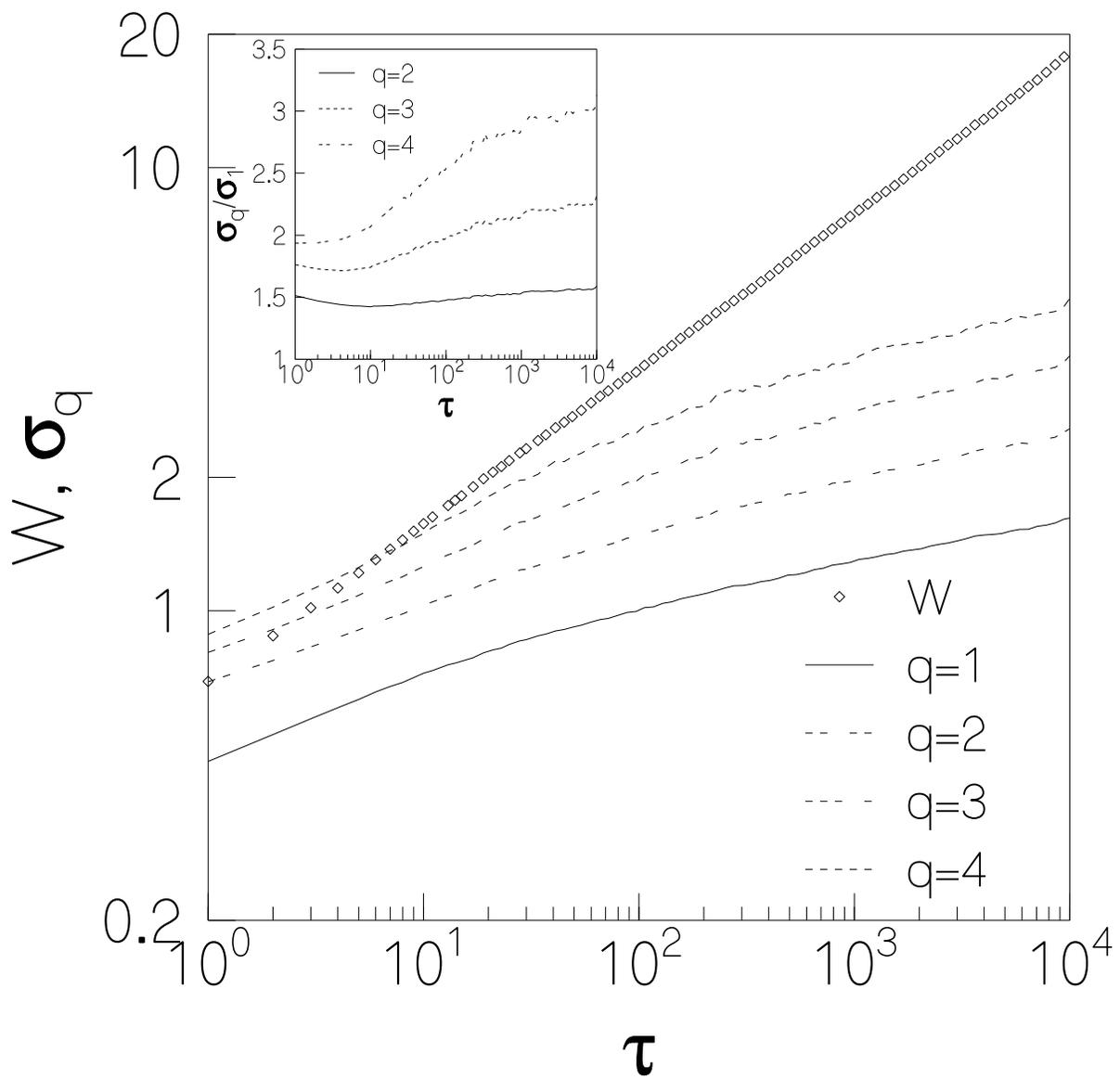
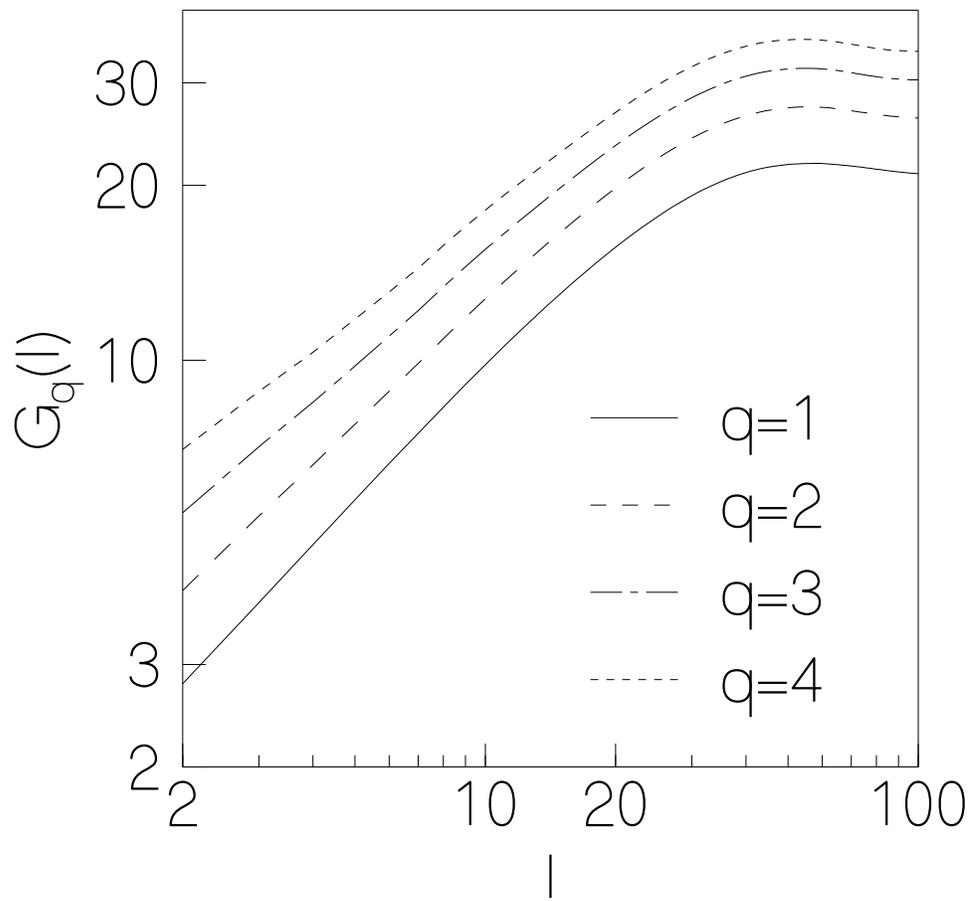
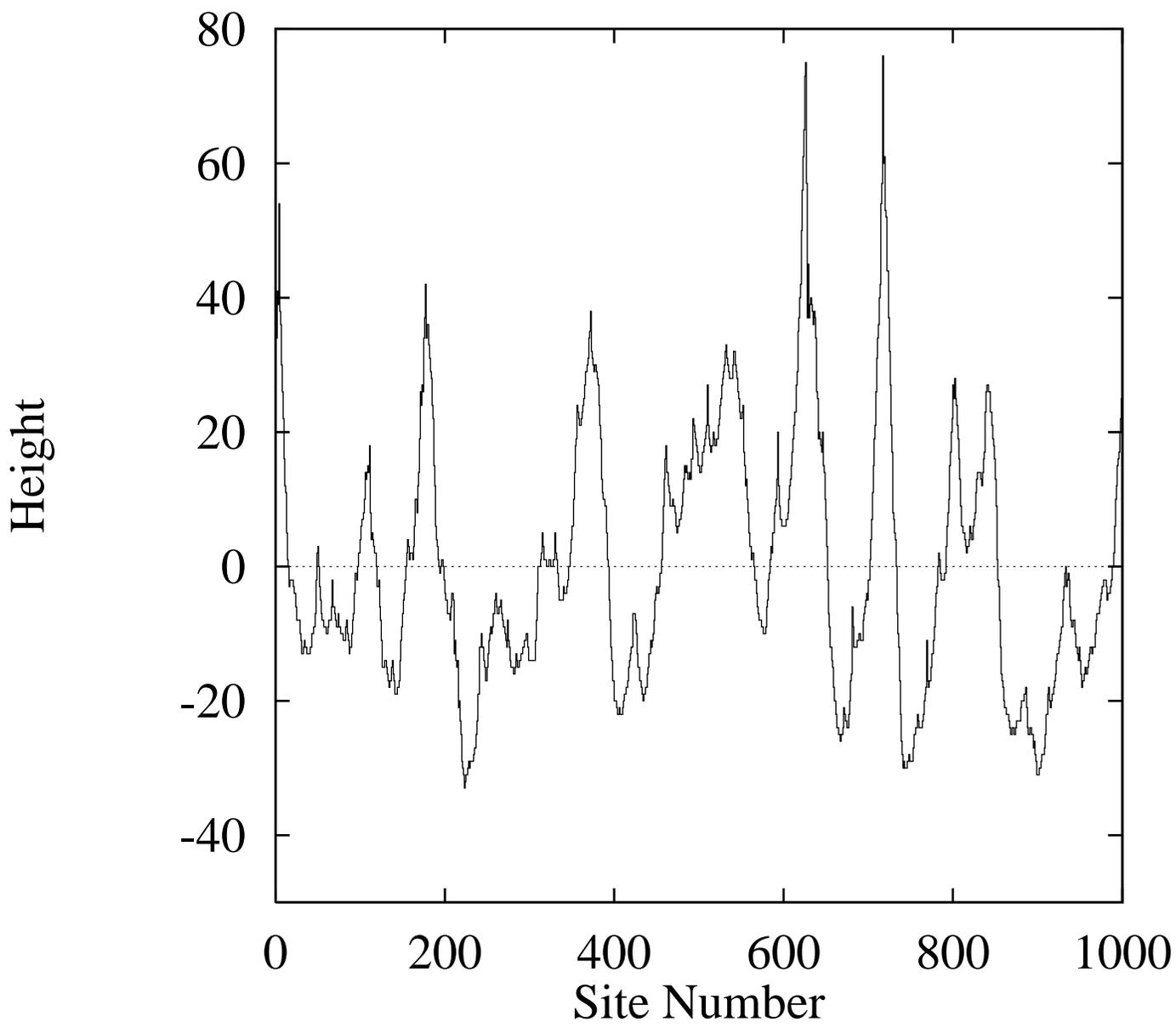
\end{document}